\documentclass[12pt]{article}
\usepackage{graphicx}
\usepackage{amsmath}
\usepackage{amssymb}
\usepackage{amsbsy}
\usepackage{bm}
\usepackage{graphics, setspace}
\newcommand{\mathsym}[1]{{}}
\newcommand{\unicode}[1]{{}}

\newcommand{\bit}{\begin{itemize}}
\newcommand{\eit}{\end{itemize}}

\def\benu{\begin{enumerate}}
\def\eenu{\end{enumerate}}
\def\noi{\noindent}
\def\btab{\begin{tabbing}}
\def\etab{\end{tabbing}}

\def\bit{\begin{itemize}}
\def\eit{\end{itemize}}
\def\beq{\begin{equation}}
\def\eeq{\end{equation}}
\def\bec{\begin{center}}
\def\eec{\end{center}}
\def\btable{\begin{tabular}}
\def\etable{\end{tabular}}
\def\beqr{\begin{eqnarray}}
\def\eeqr{\end{eqnarray}}

\def\Rarw{\Rightarrow}
\def\longrarw{\longrightarrow}

\def\om{\omega}
\def\gm{\gamma}

\def\lm{\lambda}
\def\eps{\epsilon}

\def\alp{\alpha}
\def\bt{\beta}

\def\dl{\delta}
\def\Dl{\Delta}
\def\sg{\sigma}
\def\Om{\Omega}
\def\del{\partial}

\def\half{\frac{1}{2}}

\def\btab{\begin{tabbing}}
\def\etab{\end{tabbing}}
\def\beqrs{\begin{eqnarray*}}
\def\eeqrs{\end{eqnarray*}}
\def\noi{\noindent}
\def\lan{\langle}
\def\ran{\rangle}

 \fontfamily{ptm}\selectfont
 \usepackage{mathptmx}           

\title{\bf Spectral Brilliance of Parametric X-rays at the FAST facility}

\author{Tanaji Sen \\ Accelerator Physics Center, FNAL, Batavia, IL 60510 \\
Todd Seiss \\ Princeton University, Princeton, NJ}

\date{}

\begin{document}

\maketitle


\begin{abstract}
We discuss the generation of parametric X-rays in the new photoinjector at the FAST (Fermilab
Accelerator Science and Technology) facility in
Fermilab. These experiments will be conducted in addition to channeling X-ray radiation
experiments. The low emittance electron beam makes this facility a promising source for 
creating brilliant X-rays. 
We discuss the theoretical model and present detailed calculations of the intensity
spectrum, energy and angular widths and spectral brilliance under different conditions.
We also report on expected results with parametric X-rays generated while under channeling
conditions. 
\end{abstract}

\section{Introduction}

Energetic charged particles traveling through a crystal can produce X-rays by several
mechanisms. Incoherent bremsstrahlung and transition radiation give rise to a continuous spectrum 
while channeling radiation (CR) and parametric X-ray (PXR) radiation produce 
quasi-monochromatic discrete X-ray spectra.
One of the main advantages of using crystals is that CR and PXR produce
hard X-rays with much lower energy electrons compared to, for example, synchrotron radiation 
produced X-rays in circular rings, . It takes a 3 GeV electron beam 
(assuming a bend field of 1 T) to generate X-rays with a critical energy of 10keV via synchrotron 
radiation while 10 MeV electrons have sufficed with channeling and parametric radiation at the 
same energy. Hard X-ray generation using crystals
and 50 MeV electrons is one of the planned set of experiments at Fermilab's L-band photoinjector 
in the FAST facility (formerly called ASTA) \cite{ASTA_ref, Mihalcea_15}, currently being 
commissioned . The major goal of these experiments is to demonstrate that such a 
photoinjector with a low emittance electron beam can serve as a model
for a brilliant compact X-ray source when scaled to a higher gradient X-band photoinjector. 

 The detailed characteristics of CR expected at FAST was discussed in \cite{Sen_14}. 
In this paper we will consider the spectral brilliance of PXR under
various conditions at FAST. The PXR mechanism was first discussed
several decades ago \cite{Ter-Mikaelian, Baryshevsky_71}, experimentally
measured first with electrons in 1985 \cite{Tomsk_85} and since then observed at many 
laboratories; several reviews are now available \cite{Rullhusen_98, Baryshevsky_05}.
PXR has also been observed from 400 GeV protons using a bent crystal at CERN's SPS accelerator
\cite{SPS_PXR}. 
The characteristics of PXR differ from CR in several ways.
In CR emission, the X-ray energy spectra is discrete at electron energies below 100 MeV and the 
frequencies depend on the particle energy, while in PXR they are independent of particle 
energy at relativistic particle speeds. PXR can be generated at large
angles from the particle's direction which differentiates it from both
CR  and bremsstrahlung making the background contribution significantly less than the signal.
The PXR spectral lines are also more monochromatic than CR, the width is at least an 
order of magnitude smaller. The disadvantage of PXR is that the photon yield is about two to three
orders of magnitude smaller than that of CR. On the other hand, PXR can be generated 
simultaneously with CR generation thus potentially allowing multiple X-ray beams with different
spectra and in different directions. The crystal requirements for PXR and CR production are
similar, namely high thermal conductivity, low photon absorption length, 
high dielectric susceptibility and large lattice spacing.  In this article we will 
consider the spectral brilliance from a PXR source under different conditions. In general,
the brilliance will be several orders of magnitude lower than that from the brightest X-ray
sources such as XFELs or inverse Compton scattering. However, compared to those
sources, a PXR source can deliver X-rays suitable for industrial and medical applications
with significantly lower cost, complexity and size. The special feature of the photoinjector
at FAST for generating brilliant X-rays using crystals is that low emittances can be generated 
by shaping the laser spot size on the cathode and even lower emittances have been obtained
with field emission (FE) nanotip cathodes \cite{Piot_14}. In addition, the photocathode and FE
cathodes operate at GHz frequencies, so the high repetition rate allows for low bunch charge
required for these low emittances. 

In Section 2, we discuss the PXR spectrum, notably the photon energy and the spectral distribution
dependence on the crystal geometry. In Section 3, we present calculations of the energy
width with contributions from geometrical effects and multiple Coulomb scattering, while in 
Section 4 we consider angular broadening and in Section 5 the spectral brilliance. Section 6 
contains specific calculations for the photoinjector at FAST; this includes cases with the use
of the present goniometer holding the crystal, the options with a new goniometer, and finally
PXR emission under channeling conditions. We conclude with a summary of results in Section 7.

\section{Characteristics of the PXR spectrum}  \label{sec: spectrum}

PXR emission occurs when the virtual photons accompanying the charged particle scatter off the 
atomic electrons in the crystal and interfere constructively along certain directions.
The incoming virtual photon's wave-vector ${\bf k}_i$ and the outgoing 
real photon's wave vector ${\bf k}_{f}$ are related by the Bragg condition
for momentum transfer
\beq
{\bf k}_{f} = {\bf k}_i + m {\bf g}
\eeq
where ${\bf g} $ is the reciprocal lattice vector of the scattering planes and $m$ is an integer.
Writing ${\bf k}_i = \om \hat{\bt}/v$, ${\bf k}_{f} = \om \hat{\Om}/(c/\sqrt{\eps})$,  and taking 
the scalar product of the above equation with $\hat{\bt}$ yields the outgoing PXR photon energy as
\beq
E    = \hbar \om = m \frac{\hbar c |{\bf g}\cdot \hat{\bt}|}
{1/\beta - \sqrt{\eps}\hat{\bt} \cdot \hat{\Om}}
\eeq
Here
${\bf v}= \hat{\bt}\bt c$ is the velocity vector of the particle, $\hat{\beta}$ is the unit
velocity vector.  For electron energies in the 
range of tens of MeV, we can approximate $\bt \approx 1$. 
$\hat{\Om}$ is the unit vector along the direction of the emitted photon and $\eps$ is the real 
part of the permittivity. 
If the first order spectrum is obtained by reflection from a plane with Miller indices
$(h,k,l)$, multiples of this frequency occur from reflections off planes with indices
 $(mh,mk,ml)$ with $m > 1$. 
The above equation can also be derived by requiring that
the phase difference between photons reflected from adjacent lattice planes be an integer
multiple of 2$\pi$. 
At X-ray energies, the frequency dependent real part of the dielectric 
function can be written as $\eps(\om)  \simeq 1 - (\om_p/\om)^2 $
The plasma frequency $\om_p$ for most crystals is in the range 10-90 eV
while X rays have keV range energies. 
Approximating $\eps \approx 1$, the first order ($m=1$) PXR energy can be written as 
\beq
E = \hbar c \frac{g \sin(\theta_B + \alp)}{2\sin^2((\theta_D + \alp)/2)}
\stackrel{\theta_D=2\theta_B, \alp=0}{\longrarw} \hbar c \frac{g}{2\sin\theta_B}
\label{eq: Ephoton}
\eeq
where $\theta_B$ is the angle of the crystal plane with the beam direction,
$\alp$ is the angle of the electron with the central electron beam direction,
and $\theta_D$ is the observation angle of the emitted radiation with the beam
direction. This shows that the photon energy is independent of beam energy and can be changed
by rotating the crystal with respect to the beam direction.

The expression for the angular intensity distribution calculated from a kinematic theory
can be written as \cite{Feranchuk_85, Nitta_91, Brenzinger_97}
\beqr
\frac{d^3 N}{d \om d\theta_x d\theta_y} & = & \frac{\alp_f}{4\pi}\frac{\om}{c}
f_{geo}(\hat{n}, \hat{v},\hat{\Om}) e^{-M(g)}
\frac{|\chi_g(\om)|^2}{\sin^2 \theta_B}
\left[ \frac{\theta_x^2 \cos^2 2\theta_B + \theta_y^2}
{[\theta_x^2 + \theta_y^2 + \theta_{ph}^2]^2} \right] \dl(\om - m \om_B)
\label{eq: diffang_spect} \\
f_{geo}(\hat{n}, \hat{v},\hat{\Om}) & = & 
L_a |\frac{\hat{n}\cdot \hat{\Om}}{\hat{n}\cdot\hat{v}}|
(1 - \exp[-\frac{t}{L_a(\hat{n}\cdot\hat{\Om})}]) \\
\!\!\!\! |\chi_g(\om)| &\!\!\! = \!\!\! & |S({\bf g})| \left[ \frac{\om_p^2}{\om^2}\frac{F(g)}{Z}\right], 
S(g) \!\!  = \!\!  \sum_j \exp[i {\bf g} \cdot {\bf r}_j], 
F(g) \!\! = \!\! \int d^3 r \rho({\bf r}) \exp[-i {\bf r \cdot g}]  \\
\theta_{ph}^2 & = & \frac{1}{\gm^2} + (\frac{\om_p}{\om})^2 
\label{eq: intensity_spect}
\eeqr
Here $\theta_x, \theta_y$ are the angular deviations from the direction of specular reflection
($\theta_D = 2\theta_B$) in the planes parallel and perpendicular respectively to the
diffraction plane,
$\alp_f$ is the fine structure constant, $L_a$ is the photon absorption length in
the crystal, $t$ is the crystal thickness. $\chi_{\bf g}$ is the frequency dependent
part of the Fourier transform of the dielectric susceptibility and depends on
the plasma frequency $\om_p$, the atomic number $Z$ of the crystal and on
$S({\bf g}), F({\bf g})$.  
$S({\bf g})$ is the crystal structure
function, the sum is evaluated over the atom location ${\bf r}$ in a unit cell.
For FCC cubic crystals like diamond and silicon, $S({\bf g}) \ne 0$ only if the
Miller indices $(h,k,l)$ of the reflection plane are either all odd or all even. If they
are all even, $h+k+l$ must be divisible by four. For an amorphous material $S({\bf g})=0$, hence
there is no PXR emission. 
$F({\bf g})$ is the atomic scattering form factor which is the Fourier transform of 
$\rho$, the density distribution of the atomic electrons. The Debye-Waller factor $e^{-M(g)}$
accounts for thermal vibrations and is close to unity at room temperature.

Among the assumptions made in deriving this result are that the beam is 
relativistic ($\gm \gg 1$), deviations of the photons from the Bragg specular
reflection angle are small ($\theta_x, \theta_y \ll 1$) and the crystal thickness $t$
is larger than the photon formation length.

The angular factor, the term in square brackets on the right hand side of
Eq. (\ref{eq: diffang_spect}) contains the dependence on the angles 
$\theta_x, \theta_y$, the beam energy $\gm$ and on the X-ray frequency $\om$. 
The two terms in the numerator of the angular factor describe the contributions from the 
two polarizations. If the observation angle $2\theta_B = \pi/2$, the horizontal contribution
vanishes and the observed spectrum is completely vertically polarized at all angles 
$\theta_x, \theta_y$. 
When projected onto one of the two angles the angular 
distribution can be either single peaked or double peaked, depending on the
value of the orthogonal angle. 
The extrema of the angular distribution in $\theta_x$ are at
$\theta_x = 0, \pm \sqrt{\theta_{ph}^2 + \theta_y^2(1 - 2\sec^2 2\theta_B)}$
while those in $\theta_y$ are at $\theta_y = 0,\pm
\sqrt{\theta_{ph}^2 + \theta_x^2(1 -  2\cos^2 2\theta_B)}$.
When there is a single peak, the maximum is at zero, while with double peaks, 
the maxima are close to $\pm \theta_{ph}$ and there is a local minimum at zero angle. 

\begin{figure}
\centering
\includegraphics[scale=0.65]{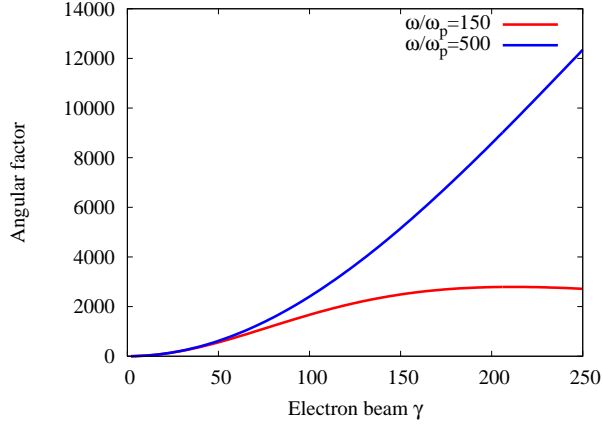}
\caption{Dependence of the angular factor with the electron beam's relativistic
factor $\gm$ for two values of $\om/\om_p$ with $\om_p=38$eV. The lower value 
$\om/\om_p$ corresponds to an X-ray energy of 5.7 keV and the higher value to a 
X-ray energy of 19keV.}
\label{fig: angfact_gam}
\end{figure}

The angular distribution depends on the beam energy only through the angle
$\theta_{ph}$ which occurs in the angular factor. At low beam energies
such that $\gm \ll (\om/\om_p)$, we have $\theta_{ph} \approx 1/\gm$ and the
intensity increases as $\gm^4$. This can also be seen by a power series 
expansion of the angular factor in terms of the parameter $\gm/(\om/\om_p)$
\beqr
\frac{\theta_x^2 \cos^2 2\theta_B + \theta_y^2}
{[\theta_x^2 + \theta_y^2 + \theta_{ph}^2]^2} & = &
\frac{(\theta_x^2 \cos^2 2\theta_B + \theta_y^2)}{(1 + \gm^2(\theta_x^2 + \theta_y^2)^2}\gm^4
\left[1 - 
\frac{2}{(1 + \gm^2(\theta_x^2 + \theta_y^2)} (\frac{\gm}{(\om/\om_p)})^2 + \right. \nonumber \\ 
&   & \left. + \frac{3}{(1 + \gm^2(\theta_x^2 + \theta_y^2)^2} (\frac{\gm}{(\om/\om_p)})^4 +
 O[(\frac{\gm}{(\om/\om_p)})^{6}] \right]
\eeqr
The denominator $(1 + \gm^2(\theta_x^2 + \theta_y^2)$ is of order unity over the useful range
of angles $\theta_x, \theta_y$.
As the beam energy increases, the angular factor and the angular intensity
distribution reaches a maximum around $\gm = \om/\om_{ph}$, levels off and then
decreases slowly at higher beam energy. This behavior can be seen in Fig.
\ref{fig: angfact_gam} in which the value of the angular factor at
 $\theta_x=0, \theta_y=1/\gm$ is plotted. 
In the lower of the two curves where $\gm$ ranges from zero to larger than $\om/\om_{p}$
the intensity levels off, 
 while in the upper curve, $\gm < \om/\om_{p}$ over the entire range, so the 
intensity grows monotonically

The atomic structure function $F(g)$ and consequently the susceptibility are calculated from the 
expressions
\beqr
F(g) & = & f_0(s) + (f^{'} + i f^{''}), \;\;\;
f_0(s)  =  \sum_i a_i \exp[-b_i s^2] + c , \;\;\;\;\;  s = 4\pi g  \nonumber \\
\chi_0 & = & -\frac{\lm^2 r_e N_C}{\pi V_C} [f^{'} + i f^{''}], \;\;\;\;
\chi_g  =  -\frac{\lm^2 r_e}{\pi V_C} [(f_0(s) + f^{'} -Z) + i f^{''} ]S_{hkl} 
\eeqr
The coefficients $(a_i, b_i, c)$ are the Cromer-Mann coefficients \cite{Cromer-Mann} while the
frequency dependent form factors $(f^{'}, f^{''})$ can be obtained from a database maintained by NIST \cite{NIST}. Here $r_e$ is the classical electron radius, $N_C$ is the number of atoms in the
unit cell, $V_C$ is the volume of the unit cell, $Z$ is the atomic number of the crystal
and $S_{hkl}$ is the crystal structure factor for the plane with indices $(h,k,l)$. The photon 
attenuation length at a wavelength $\lm$ can be found from $L_a = \lm/(2\pi |{\rm Im}(\chi_0)|)$.

\begin{figure}
\centering
\includegraphics[scale=0.45]{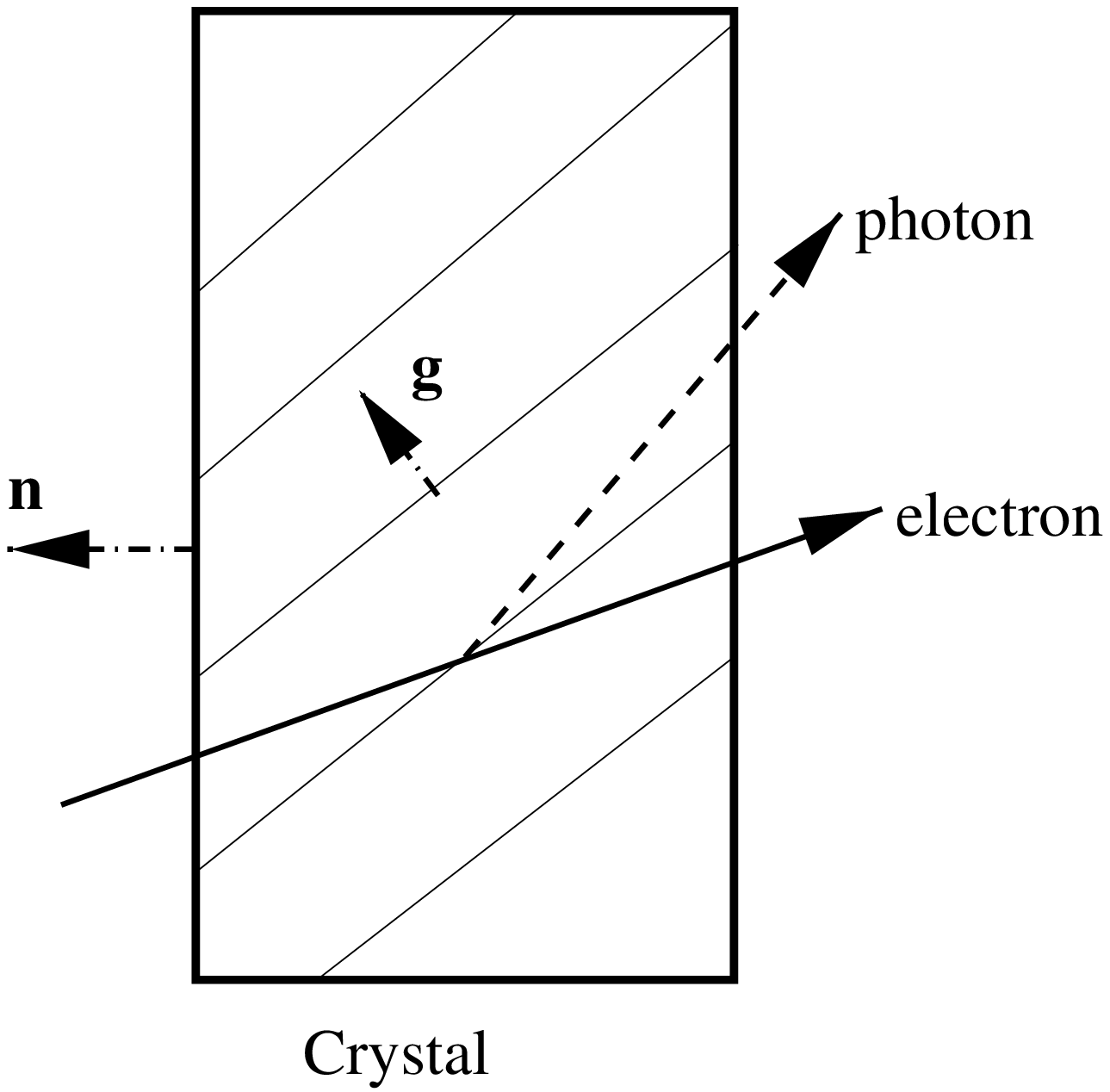}
\includegraphics[scale=0.45]{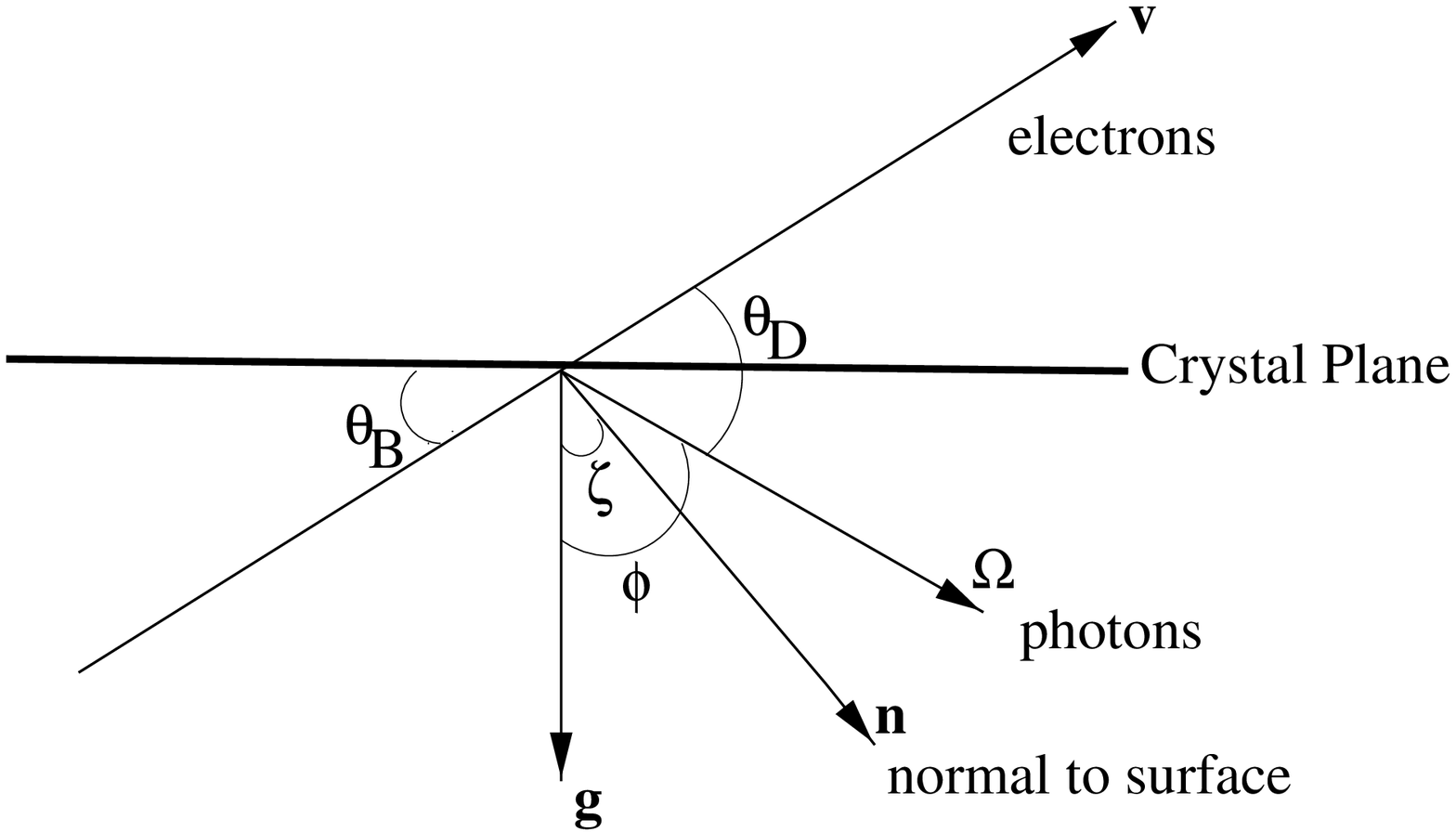}
\caption{Left: A general crystal geometry for an arbitrary orientation of the reflecting plane.
Reciprocal lattice vector ${\bf g}$: normal to the crystal plane, {\bf n}: normal to the
crystal surface. Bragg geometry corresponds to the crystal plane which has ${\bf g} || {\bf n}$,
Laue geometry to ${\bf g} \perp {\bf n}$. Right: The different angles between 
${\bf g}$, ${\bf n}$, ${\bf v}$: electron velocity vector, and
${\Om}$: direction of the detector's central axis}
\label{fig: geometries}
\end{figure}

We now discuss the geometric factor $f_{geo}$ in Eq.(\ref{eq: diffang_spect}).
The unit vectors $\hat{n}, \hat{v},\hat{\Om}$ are defined and shown in 
Fig.\ref{fig: geometries}. 
The figure on the left in Figure \ref{fig: geometries} shows the electron velocity
vector {\bf v}, the  normal ${\bf n}$ to the crystal surface, 
and the normal ${\bf g}$ to the crystal planes. 
The figure on the right in Figure \ref{fig: geometries} shows 
the direction of the detector's central axis $\bm{\Om}$, and the angles between the vectors.
It is clear that in order for the
crystal plane to be a reflecting plane, the angle $\pi/2 - \theta_B$ between 
${\bf g}$ and {\bf -v} and also the angle $\phi$ between ${\bf g}$ and
$\bm{\Om}$ must satisfy $0 \le (\pi/2-\theta_B, \phi, \zeta) \le \pi/2 $
where $\zeta$ is the angle between ${\bf g}$and {\bf n}. Bragg geometry corresponds to
$\zeta=0$ while in Laue geometry $\zeta=\pi/2$. 
If $\theta_D$ is the angle of the detector relative to the velocity vector 
{\bf v}, then we have
\[ \phi = \frac{\pi}{2} - (\theta_D - \theta_B), \;\;\;
\hat{n}\cdot\hat{v} = -\sin(\theta_B + \zeta), \;\;\;\; 
\hat{n}\cdot\hat{\Om} = \sin(\theta_D - \theta_B - \zeta)
\]
The geometric factor $f_{geo}$ in the intensity expression can then be written as
\beq
f_{geo}  =  L_a |\frac{\sin(\theta_D - \theta_B - \zeta)}{\sin(\theta_B + \zeta)}|
[ 1 - \exp(- \frac{t}{L_a|\sin(\theta_D - \theta_B - \zeta)|}) ]
\eeq
Note that if $\theta_D=\theta_B + \zeta$, the geometric factor $f_{geo}=0$ because
the photons travel along the larger transverse dimensions of the crystal and
will be mostly absorbed in the crystal.

We mention here that a refinement to the kinematical theory is the dynamical diffraction theory
\cite{Baryshevsky_72, Garibian_72, Caticha_89} which takes into account the coupling between the
photon fields with wave vectors ${\bf k_i}$ and ${\bf k_f + g}$ via interaction with the crystal.
This coupling gives rise to additional PXR photons emitted in the forward direction in 
close proximity to the electron beam. 
This forward PXR was observed in experiments at the Mainz laboratory \cite{Backe_05} but care was
required to extract this PXR emission signal as transition radiation and bremsstrahlung are also 
emitted in the same direction. We will not
discuss this forward PXR emission here as it does not offer the relatively background free
property of PXR emission at the Bragg angle.

\section{Energy spectrum broadening}

The intensity spectrum given by Eq.(\ref{eq: intensity_spect}) predicts a delta function spectrum
at integer multiples $m\om_B$ of the Bragg frequency. In practice, there are 
several mechanisms which broaden the frequency of each line in the spectral distribution. We 
discuss the important sources and present analytical results for their contributions to the
energy width and compare them to previous experimental results. 

\subsection{Geometric effects}

\begin{figure}
\centering
\includegraphics[scale=0.65]{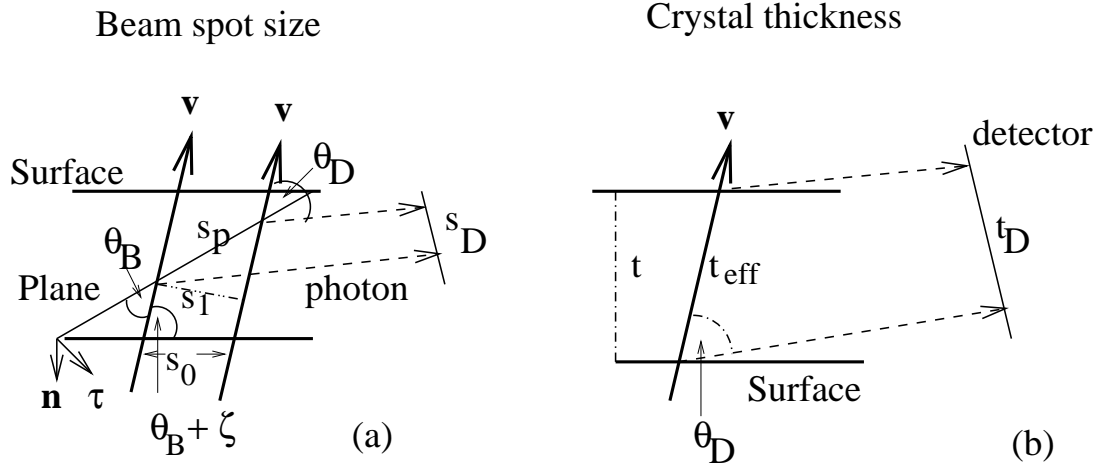}
\caption{Effect of the beam spot size (a) and crystal thickness (b) on the 
angular photon spread. In (a), the beam spot diameter is represented by $s_0$
and $s_D$ is its projection on the detector. In (b), the crystal thickness is
$t$, the effective thickness traversed by the electrons is $t_{eff}$ and 
$t_D$ the projection on the detector.}
\label{fig: Ewidth}
\end{figure}

From Eq.(\ref{eq: Ephoton}), it follows
that the photon energy depends on the angle of the incident electron and
the photon direction. The incident angle will have a finite spread due to the 
beam divergence while the photon angle hitting the detector can have a spread
due to several effects including the finite beam spot size on the crystal
surface, the detector size and the crystal thickness. 

Writing the spread in electron incident angle in the diffraction plane as 
$\Dl\alp_x$ and the photon angle as $\theta_D = 2\theta_B + \theta_x$, we have 
from Eq.(\ref{eq: Ephoton}) to first order in $\Dl \alp_x$
\beq
(\frac{\Dl E}{E})_{div} = \frac{1}{2\sin^2 \theta_B}\theta_x \Dl\alp_x
\eeq
We set $\Dl\alp_x = \sg_x^{'}$ the beam divergence in the diffraction plane
and $\theta_x = \Dl x_{det}/R$ where $\Dl x_{det}$ is the width of the detector
and $R$ is the distance from the crystal to the detector.

Again from Eq.(\ref{eq: Ephoton}), it follows that the energy spread is
related to the photon angular spread as
\beq
(\frac{\Dl E}{E})_{D} = \frac{\Dl \theta_D}{\tan\theta_B}
\eeq
The impact of the finite beam size on $\Dl\theta_D$ can be seen in 
Fig.\ref{fig: Ewidth}a. A finite
size on the crystal surface projects to a spot size on the reflecting planes
from which photons with a spread of angles can reach the detector. Let $s_0$ be
the beam spot diameter on the crystal and $s_p$ its projection on the crystal's
reflecting plane. The angle the velocity vector $\hat{v}$ makes with the 
crystal surface is $\theta_B+\zeta$, hence it follows from the figure that
\[ s_1 = s_0\sin(\theta_B+\zeta), \;\;\; s_p = \frac{s_1}{\cos(\pi/2-\theta_B)},
\;\;\; \Rarw s_p = \frac{\sin(\theta_B+\zeta)}{\sin{\theta_B}}s_0 \]
On reflection from the crystal plane, the size $s_p$ projects to a size
$s_D = s_p\sin(\theta_D-\theta_B)$. The angular spread in the photon angles
resulting from the beam size is 
\beq
(\Dl\theta_D)_{size} = \frac{s_D}{R} = \frac{\sg_x}{R}
\frac{\sin{\theta_B+\zeta}}{\sin\theta_B} \sin(\theta_D-\theta_B)
\eeq
where we have replaced $s_0$ by the  rms beam size $\sg_x$. 

The impact of the finite crystal thickness is seen in Fig.\ref{fig: Ewidth}b.
The effective crystal thickness $t_{eff}$ projects to a length on the
detector $t_D = t_{eff}\sin\theta_D = t\sin\theta_D/\sin(\theta_B+\zeta)$
where $t$ is the crystal thickness. Hence the angular spread in photon angles
due to the crystal thickness is
\beq
(\Dl\theta_D)_{crystal} = \frac{t_D}{R} = \frac{t}{R}
\frac{\sin \theta_D}{\sin(\theta_B+\zeta)}
\eeq
Finally the detector size results in an angular spread 
$(\Dl\theta_D)_{det} = \Dl x_{det}/R$.

Adding the independent sources of beam angle spread and photon angle spread in
quadrature, the energy spread due to these geometric effects is
\beq
\frac{\Dl E}{E} = \frac{1}{R}\left[(\frac{\Dl x_{det} \sg_x^{'}}{2\sin^2 \theta_B})^2
+ \cot^2\theta_B\left\{ (\sg_x\sin(\theta_B+\zeta))^2 + 
(t \frac{\sin 2\theta_B}{\sin(\theta_B+\zeta)})^2
+ (\Dl x_{det})^2\right\} \right]^{1/2}
\label{eq: Ewidth_geom}
\eeq
Here we have set $\theta_D=2\theta_B$, the direction for specular reflection.
For typical beam and crystal parameters, the dominant contributions are from
the beam spot size and the detector width while the contributions from the beam
divergence and crystal thickness are significantly smaller.

\subsection{Multiple Scattering}

The contribution of multiple Coulomb scattering can be analytically calculated from the
differential angular spectrum. Using Eq.(\ref{eq: diffang_spect}), it follows
that the differential angular spectrum per unit length is given by 
\beq
\frac{d^3 N}{dz\; d\theta_x\; d\theta_y} = 
\frac{\alp_f}{4\pi c}\frac{\om_B}{\sin^2\theta_B} 
\frac{1}{|\hat{n}\cdot\hat{v}|}\exp[- \frac{z}{L_a|\hat{n}\cdot \hat{\Om}}]
e^{-M(g)} |\chi_g(\om)|^2
\frac{\theta_x^2 \cos^2 2\theta_B + \theta_y^2}
{[\theta_x^2 + \theta_y^2 + \theta_{ph}^2]^2} 
\eeq
The spread in angles $\theta_x, \theta_y$ changes as the electron beam propagates
through the crystal due to multiple scattering. Assuming the multiple scattering
process to be Gaussian, the angles are sampled from distributions
\[ f(\theta_x) = \frac{1}{\sqrt{2\pi}\sg_x^{'}}
\exp[-\frac{\theta_x^2}{2\sg_x^{'2}}] \] 
and a similar expression for the distribution in $\theta_y$. Here 
$\sg_x^{'},\sg_y^{'}$ are the rms beam divergences which increase with $z$ as
the beam propagates through the crystal. Writing the initial beam divergences as
$(\sg_{x,0}^{'}, \sg_{y,0}^{'})$, the $z$ dependent divergences are
\[ \sg_x^{'}(z) = [(\sg_{x,0}^{'})^2 + (\sg_{MS}^{'}(z))^2]^{1/2} \]
and similarly for $\sg_y^{'}$. Here $\sg_{x,MS}$ is the rms multiple scattering
angle in the $x$ direction. This can be found from the expression \cite{PDG}
\beq
\sg_{MS}^{'}(z) = \frac{13.6}{E_e}\sqrt{\frac{z}{X_{rad}}}[ 1 + 0.038\log(\frac{z}{X_{rad}})]
\eeq
where $z$ is the path length traversed in the crystal,$E_e$ is the electron beam energy in MeV
and $X_{rad}$ is the radiation length. 

The multiple scattering weighted distribution function in $\theta_x$ is then
\beqr
(\frac{dN}{d\theta_x})_{MS} &  = &  \int dz \int d\phi_x \int d\theta_y 
f(\phi_x) \frac{d^3N}{dz d\theta_x d\theta_y}(z,\phi_x,\theta_x\theta_y) 
\nonumber \\
& = &  {\cal A}  \int dz \int d\phi_x \int d\theta_y 
\frac{1}{E(\theta_x+\phi_x)}\frac{1}{\hat{n}\cdot\hat{v}}
\exp[- \frac{z}{L_a|\hat{n}\cdot \hat{\Om}}]  \nonumber \\
&  & \frac{1}{\sg_x^{'}(z)}\exp[-\frac{\phi_x^2}{2\sg_x^{'2}}]
\frac{\theta_x^2 \cos^2 2\theta_B + \theta_y^2}
{[\theta_x^2 + \theta_{y}^2 + \theta_{ph}^2]^2} 
\label{eq: Ewidth_MS}
\eeqr
Here ${\cal A}$ includes all factors which do not depend on $z,\theta_x, \theta_y$.
Within the integrand, $E(\theta_x+\phi_x)$ denotes that the 
photon energy is evaluated at the angle $(\theta_x+\phi_x)$.
From this weighted distribution, the average and rms width of the energy spectrum
can be found as
\beqr
\lan E \ran & = & \frac{\int E dN}{\int dN} = \frac{\int E (dN/d\theta_x)_{MS}
  d\theta_x}{\int (dN/d\theta_x)_{MS}d\theta_x} \nonumber   \\ 
 \sg_{E, MS}^2 &  = &  \lan E^2 \ran - \lan E \ran^2  \label{eq: rms_Ewidth_MS}
\eeqr
We will use Eq.(\ref{eq: rms_Ewidth_MS}) to estimate the energy width due to
multiple scattering instead of a Monte-Carlo simulation that is often used.

There is another contribution to the linewidth from the photo-absorption in the 
crystal.
Assuming a point source electron beam and no imperfections or multiple scattering,
the photon wave train emitted by the beam has an intrinsic energy width given by \cite{Caticha_92}
 \beq 
 \Dl E_{Intrinsic} = \hbar\om \frac{\chi_{0,I}}{2 \sin^2\theta_B} = 
\frac{\hbar c}{2\sin^2\theta_B L_a}
 \eeq
 where $\chi_{0,I}$ is the imaginary part of the mean dielectric susceptibility.
The intrinsic width has a minimum value for backward emission when $\theta_B = \pi/2$. 
This contribution is typically orders of magnitude smaller than the other contributions discussed
above.

The expressions for the energy width have been checked against measured values
from a couple of earlier experiments, one at a low beam energy of 6.8MeV 
\cite{Freudenberger} and
the other with beam energy of 56 MeV \cite{Sones_thesis},  close to the FAST 
beam 
energy of 50 MeV. Since multiple 
scattering is more important at lower energy, comparison with the low
energy result are a good check of the width from multiple scattering
while the second case will be a good check of the geometrical width.
Table \ref{table: linewidth_comp} shows the results of the comparison.
 \begin{table}
 \bec
 \btable{|c|c|c|c|c|c|} \hline
 Crystal & Geometrical & Multiple Scattering & Intrinsic & Theory(total) & Experiment \\ 
 & [eV] & [eV] & [eV] & [eV] & [eV]  \\ \hline
 C (111) &  35.9 & 34.8 & 2.1$\times 10^{-4}$ &  49.4 & 51 \\
 \hline
 Si (220) & 524.9 & 61.3 & 2.8$\times 10^{-4}$ & 529 &  540 $\pm$ 120 \\
 Si (400) & 97.1 & 9.7  &  2.1$\times 10^{-3}$ & 98 & 134 $\pm$ 56 \\
 \hline
 \etable
 \eec
 \caption{Comparison of theoretical energy widths with experimental values after removing the 
effects of the detector energy resolution. 
 First row: Experiments with 6.8 MeV electrons and a diamond crystal \cite{Freudenberger}.
Second and third rows: Experiments with 50-60 MeV electrons and silicon crystals
 \cite{Sones_thesis, Sones_06}. These values are taken from Table 36 in \cite{Sones_thesis}. 
The theoretical estimates include geometrical, multiple scattering and intrinsic contributions 
added in quadrature.}
\label{table: linewidth_comp}
\end{table}
The geometrical linewidth was calculated using Eq.(\ref{eq: Ewidth_geom}) and the
multiple scattering contribution using Eqs.(\ref{eq: Ewidth_MS}) and 
(\ref{eq: rms_Ewidth_MS}). The two were then added in
quadrature to yield the theoretical value shown in Table 
\ref{table: linewidth_comp}.

\section{Angular spectrum broadening}

The measured angular intensity distribution represents a convolution of the intrinsic
PXR intensity with the Gaussian response of the detector angular 
resolution, the beam divergence and multiple scattering. Hence
the measured intensity is of the form
\beq
\frac{d^2 N}{d\theta_x\theta_y}_{conv}(\theta_x, \theta_y) = 
\frac{A_C}{2\pi \sg_x^T \sg_y^T}
\int \frac{d^2 N}{d\theta_x\theta_y}(\theta_x - q_x,\theta_y - q_y) 
\exp[-(\frac{q_x}{\sqrt{2}\sg_x^T})^2 - (\frac{q_y}{\sqrt{2}\sg_y^T})^2]
dq_x dq_y
\eeq
where $A_C$ is a constant to ensure photon number conservation after the
convolution, 
$d^2 N/d\theta_x\theta_y$ is the  angular distribution without convolution,
and the total angular resolutions $\sg_x^T,\sg_y^T$ are given by
\beq
\sg_u^T = \sqrt{(\frac{\sg_{D,u}}{R})^2 + (\sg_u^{'})^2 + (\sg_{MS}')^2}
\eeq
where $u=(x,y)$, $\sg_{D,u}$ is the detector resolution, $R$ is the distance
of the detector from the crystal, $\sg_u^{'}$ is the beam divergence
at the crystal, $\sg_{MS}^{'}$ is the effective multiple scattering angle
averaged over the path length $t/\cos\theta_B$. 

In the absence of the broadening due to the convolution, the
intrinsic angular width of he PXR spectral distribution is given by
$\Dl \theta_{PXR} \simeq \theta_{ph}$. 
In crystals with thicknesses comparable to the attenuation length, the
broadening due to multiple scattering is significant and the characteristic
double peaked angular distribution is replaced by a broadened single
peak distribution with the center filled in. 

Other sources of angular broadening are the higher order reflections from planes with spacings
which are integer sub-multiples of the primary plane. These higher order reflections produce
higher photon energies with lower yield leads to a broadened distribution when recorded on a 
detector which sums over all photon energies. 

\section{Spectral Brilliance}

The spectral brilliance of the photon beam is defined as the number of photons
emitted per second per unit area of the photon beam per unit solid angle per
unit relative bandwidth
\beq
B = \frac{d^4 N}{dt dA d\Om d\om/\om}
\eeq
Expressed in conventional light source units, the average spectral brilliance can be written in 
terms of the averaged beam parameters and differential angular intensity spectrum per electron in
a 0.1\% bandwidth
\beqr
B_{av} &  = &  \frac{d^2 N}{\hbar d\om d\Om}\frac{I_{av}}{e}\frac{E_{\gm}}{(\sg_{\gm})^2}
 *10^{-3} \\
& = & \frac{I_{av}}{e} \frac{1 }{\Dl E_{\gm}/E_{\gm}}
\frac{dN}{d\Om} \lan \frac{1}{\sg_e^2} \ran \times  10^{-15}
 \;\;\; {\rm photons/s-(mm-mrad)^2- 0.1\% BW}
\label{eq: Brill_av}
\eeqr
where $I_{av}$ is the average electron beam current, $E_{\gm}$ is the energy
of the X-ray line and $\sg_{\gm}$ is the X-ray beam spot size. In the second line
$dN/d\Om$ is the angular yield in units of photons/(el-sr), and
we set the photon spot size to the electron beam spot size in the crystal, i.e.
 $\sg_{\gm} = \sg_e $. Here $\Dl E_{\gm}$ includes only the contributions to the spectral width 
from the crystal but not that from the detector resolution. 

Due to multiple scattering within the crystal, the electron beam divergence, beam size and 
emittance will grow as the electrons move through the crystal. Writing $\eps_N$ as the normalized
electron  emittance, the emittance growth as a function of the distance traversed $z$ is 
\beq
\Dl\eps_{N,x}(z) = \gm \bt_x (\sg_{MS}^{'}(z))^2
\eeq
where $\bt_x$ is the optical beta function at the crystal in the $x$ axis, If the initial
beam size at the crystal is $\sg_{x,0}$ and the normalized emittance at the crystal entrance is
$\eps_{N,0}$, then $\bt_x = \gm(\sg_{x,0}^2/\eps_{N,0}$. The beam divergence grows as
$ \sg_e^{'}(z) = \sqrt{(\sg_e^{'}(0))^2 + (\sg_{MS}^{'}(z))^2}$
and the average of the inverse beam size squared follows from
\beq
\lan \frac{1}{\sg_e^2} \ran \equiv 
\frac{\gm}{\bt_x}\lan \frac{1}{\eps_N(z)} \ran = 
\frac{\gm}{\bt_x t_{eff}}\int_0^{t_{eff}}dz \; \frac{1}{[\eps_{N,0} +
\gm \bt_x (\sg_{MS}^{'}(z))^2]}
\eeq
This averaged expression will be used in Eq.(\ref{eq: Brill_av}) for the average brilliance.
If the initial beam divergence and beam emittance are small compared to their increase through
the crystal, and neglecting the logarithmic correction to the multiple scattering angle
, we have $(\sg_e(z),\sg_e^{'}(z) \propto \sqrt{z}$ and the emittance
grows as $\eps_N(z) \propto z$. 

Since the brilliance scales linearly with the electron current but inversely as the
square of the electron spot size, it is advantageous to geneate as small a spot size
as feasible even at the cost of reducing the beam current. Since the emittance  grows 
with crystal thickness faster than the yield does, the maximum brilliance will also require the
use of thin crystals, as will be shown later with numerical examples.

\section{PXR in FAST }

\begin{table}
\bec
\btable{|ccc|} \hline
Parameter & Value & Units \\ \hline
Beam energy & 50 &  MeV \\
Bunch charge & 20 & pC  \\
Length of a macropulse & 1 & ms \\
Number of bunches/macropulse & 2000 & - \\
Macropulse repetition rate & 5 & Hz \\
Bunch frequency & 2 &  MHz \\
Interval between bunches & 0.33 & $\mu$s \\
Bunch length & 3 &  ps \\
Crystal, thickness & Diamond, 168&  $\mu$m \\
\hline
\etable
\eec
\caption{Bunch and macropulse parameters in FAST}
\label{table: FAST_pulse}
\end{table}

In the FAST beamline, a goniometer on loan from the HZDR facility, described in \cite{Wagner},
is presently available. It has two ports through which radiation can be extracted - one 
along the beam axis which will be used for channeling radiation and another at 90$^{\circ}$
to the beam axis which can be used to extract PXR. 
This determines that with the detector angle at 90 degrees, the Bragg angle must
be 45 degrees in order to generate PXR with sufficient intensity. 

The goniometer already has a diamond crystal inside with its surface cut parallel to the (1,1,0)
 plane to generate channeling radiation from this plane. It has a thickness of 168$\mu$m and
this will be the assumed value for most calculations reported here. In order to limit heating 
the crystal by the beam, we will assume low current operation with an average beam current of
200nA and a bunch charge of 20pC. At such charge values, low transverse emittance of the order
of 100 nm can be obtained by suitably shaping the laser spot on the cathode \cite{Li_12} or
alternatively with field emission cathodes \cite{Piot_14}. The main parameters of FAST and
the crystal are shown in Table \ref{table: FAST_pulse}.
We choose $R=1$ m as the crystal to detector distance and the active size of the detector plate to
be 2cm x 2cm. 

Table \ref{table: FAST_yields_present} shows the PXR photon energies, the angular yield and the
energy width  from reflection off three of the possible low order planes with a 50 MeV electron 
beam. 
The yields include the effect of attenuation in air from the crystal to the detector. 
The 
reciprocal lattice spacing $g$ between adjacent planes with indices $(h,k,l)$ is found from
$g = (2\pi/a)\sqrt{h^2+k^2+l^2}$ where $a$ is the length of a unit cell. Consequently
both the energy and absolute linewidths increase with increasing order. The yields are higher
for the (2,2,0) and the (4,0,0) planes primarily due to the higher susceptibility $\chi_{\bf g}$.
\begin{table}
\bec
\btable{|c|c|c|c|c|c|c|} \hline
Plane & X-ray energy  & $L_{a,C}$ & $L_{a,air}$ & Attenuation  & Yield & 
$\Dl E$ \\ 
&   [keV] & [cm] & [cm]  & in air  & [photons/el-sr] & [eV] \\ \hline
(1,1,1) & 4.26 & 0.0097 & 12.72 & 3.8$\times 10^{-4}$ &  3.7$\times 10^{-7}$ & 
59 \\ 
(2,2,0) & 6.95 & 0.043 & 57.2 & 0.17 &  9.9$\times 10^{-5}$ &  93  \\
(4,0,0) & 9.83 & 0.120 & 144.9 & 0.50   & 8.8$\times 10^{-5}$  & 131 \\
\hline
\etable
\eec
\caption{Photon yields and linewidths at a Bragg angle of 45$^{\circ}$,
observation angle of 90$^{\circ}$ in FAST from PXR off some planes. Bragg geometry in 
all cases, crystal thickness = 0.168mm. The relative energy width in all cases is
about 1\%.  The yield value includes the effect of attenuation over a 1m long 
path in air from the crystal to the detector.}
\label{table: FAST_yields_present}
\end{table}
Table \ref{table: FAST_widths} shows linewidth contributions from geometric effects and 
multiple scattering for each of the planes. The two effects are comparable in
these cases. 
\begin{table}
\bec
\btable{|c|c|c|c|} \hline
Plane & Geometrical Width (eV) & Multiple-Scattering(eV) & Total (eV) \\ \hline
(111)  & 42.8 & 39.3 &  59.1 \\
(220)  & 69.5 & 62.0 &  93.2 \\
(400)  & 98.3 & 86.7 & 131.1  \\
\hline
\etable
\eec
\caption{Contributions to the energy width from geometric effects and multiple scattering.}
\label{table: FAST_widths}
\end{table}

\begin{figure}
\centering
\includegraphics[scale=0.35]{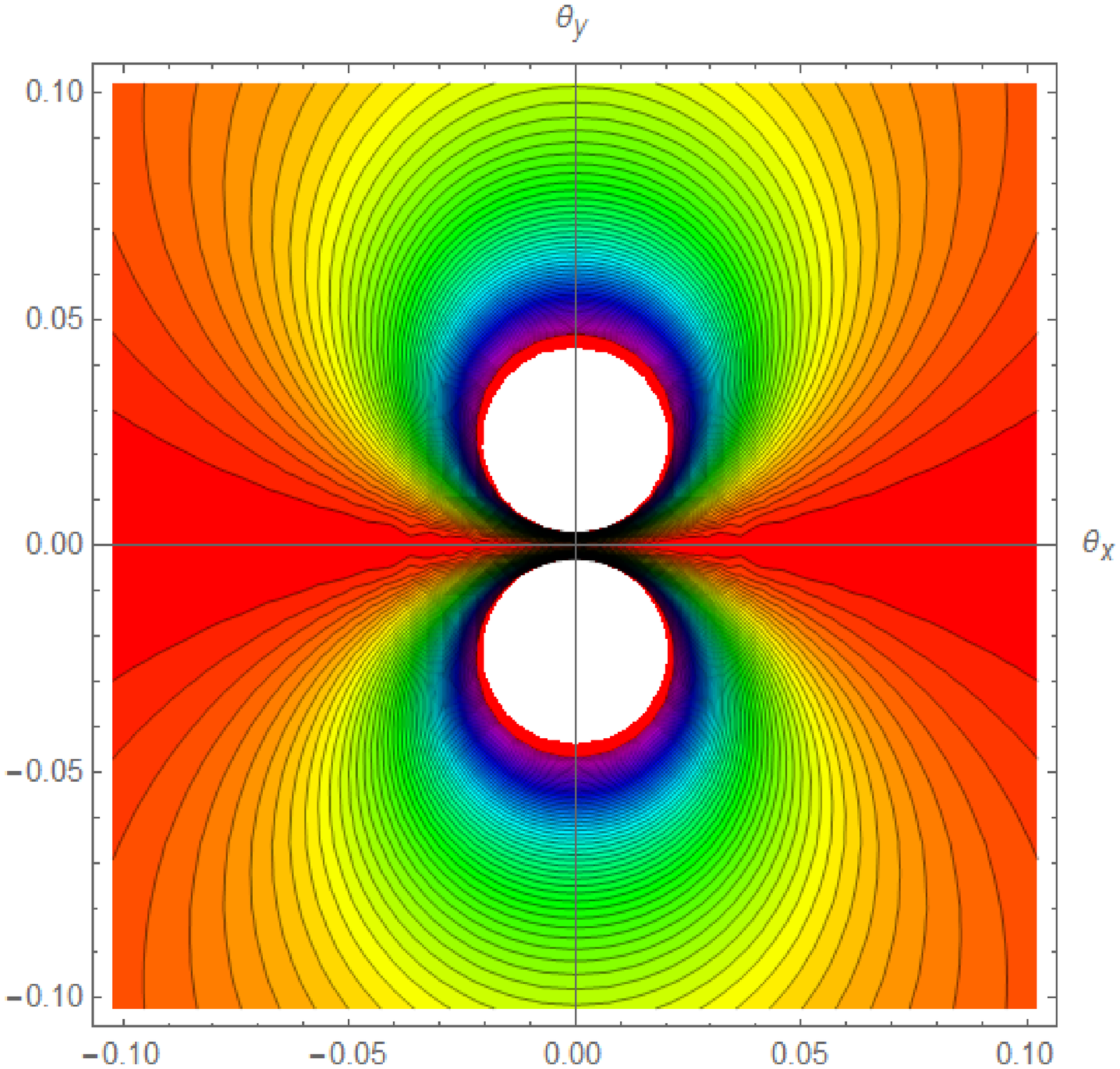}
\includegraphics[scale=0.35]{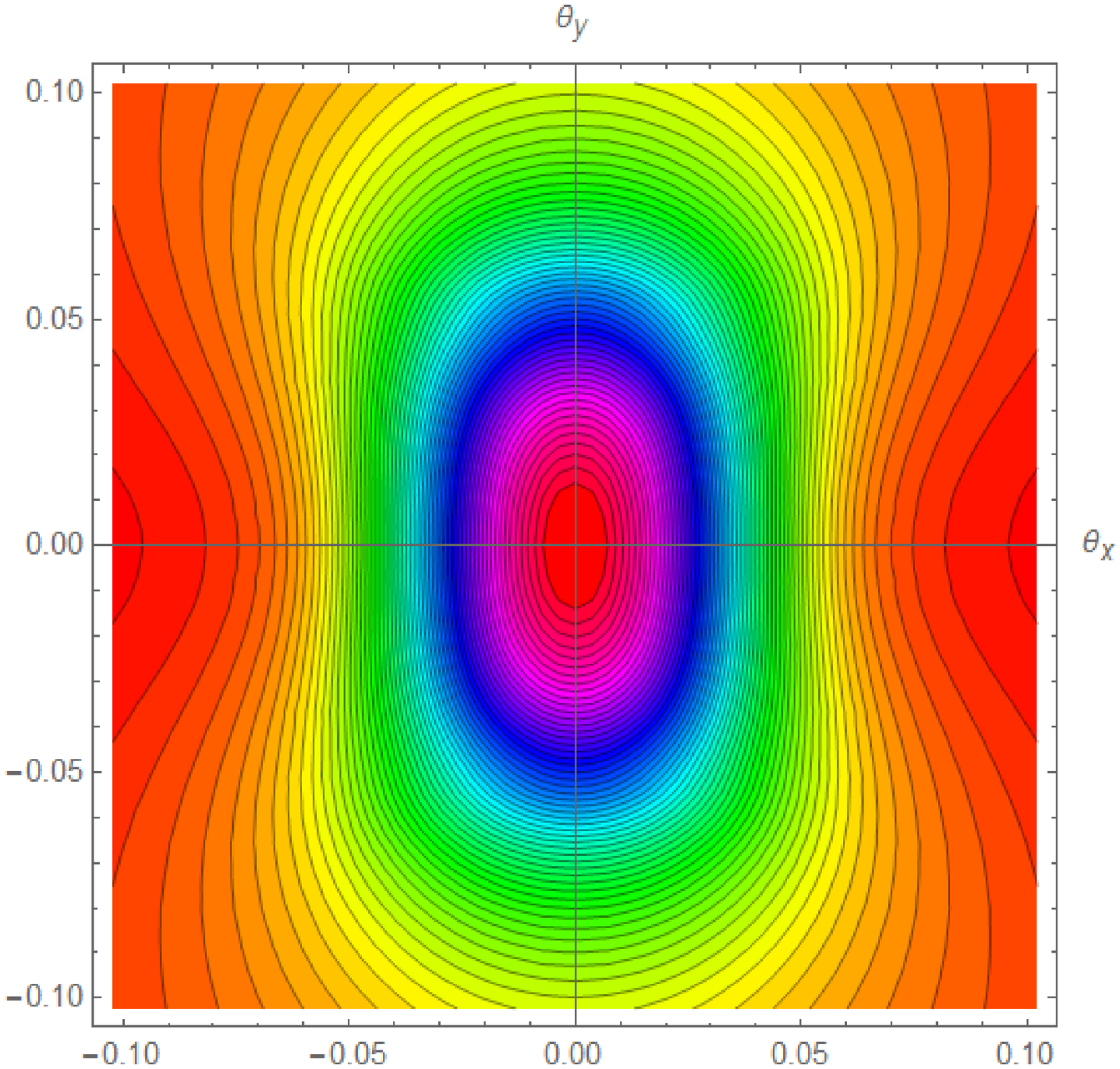}
\caption{Contour plots of the differential yields as functions of $(\theta_x, \theta_y)$ 
without broadening (left) and with broadening(right). The white areas in the left figure 
correspond to the highest intensities. The convolution lowers the peak intensities and results in
a central broad peak instead of the two peaks without convolution. Both figures are for the 
(2, 2, 0) plane with a
Bragg angle of 45$^{\circ}$ and a beam divergence of 1mrad. These contour plots correspond to the
images that would be observed on a detector plate. }
\label{fig: difyields_202}
\end{figure}

Figure \ref{fig: difyields_202} shows the two dimensional contour plots of the angular
intensity spectrum projected on the $(\theta_x, \theta_y)$ axes without and with convolution.
The broadening effects are clearly visible as the two distinct peaks in the left figure
merge into a single wider maxima in the right figure. 

For each primary plane, PXR emission also occurs from higher order planes at higher energies
and lower intensities. 
For the (1,1,1) plane, the next allowed higher order plane 
is the (3,3,3) plane, since $S(g) = 0$ for the (2,2,2)  plane. The photon energy from the
(3,3,3) plane is 12.8 keV with significantly reduced attenuation in air and resulting in a
angular yield at the detector about two orders of magnitude higher than that from the (1,1,1) 
plane. 
For the (220) plane, second order reflections from (440) are allowed with photon
energy of 13.9keV. The (440) plane has an angular yield about a third smaller than the (220) 
plane even after including smaller attenuation at the higher energy. 
For the (400) plane, the second order reflection from the (800) plane produces 19.7keV  photons
and an angular yield about 10\% that of the first order yield.
The broadening of the angular distributions from the higher order reflections in these cases is 
found to be insignificant.

\begin{figure}
\centering
\includegraphics[scale=0.55]{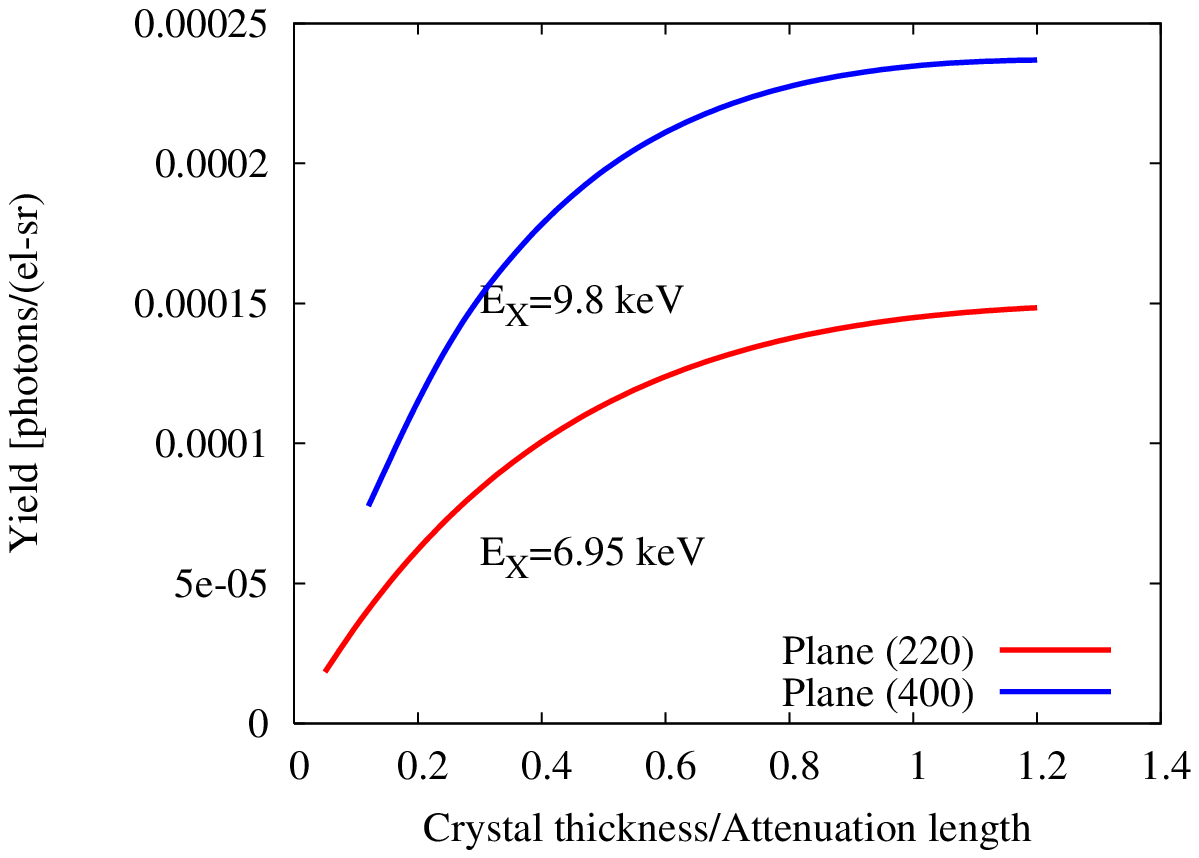}
\includegraphics[scale=0.55]{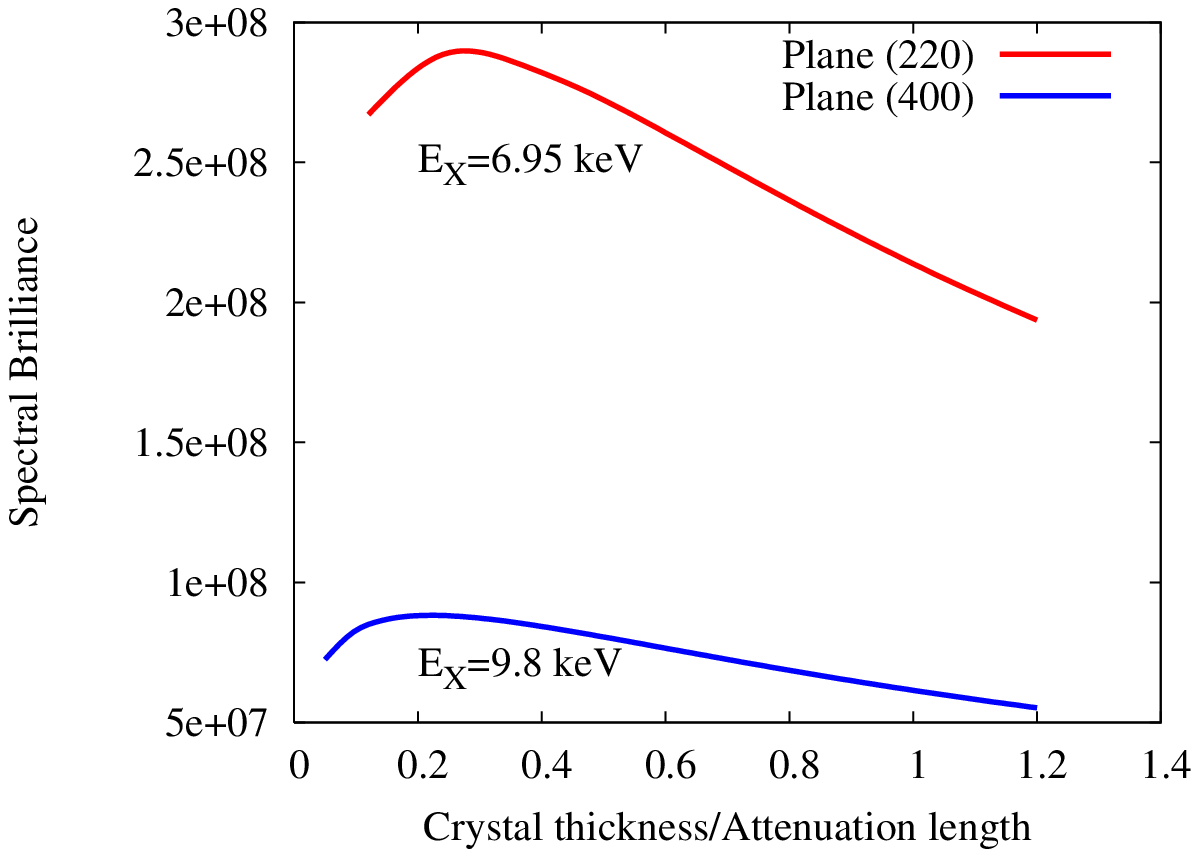}
\caption{Left: Yield per electron per steradian as a function of the crystal thickness
relative to the photon attenuation length in the crystal for two different
planes: (220) and (400). Photon attenuation lengths in the crystal are: 431 $\mu$m at 6.9keV, 
and 1198$\mu$m at 9.8keV. 
Right: Spectral brilliance as a function of the relative crystal thickness for the two planes.}
\label{fig: yield_sr_Brilliance_LC}
\end{figure}
Figure \ref{fig: yield_sr_Brilliance_LC} shows the impact of the crystal thickness on the
angular yield and spectral brilliance for two different planes. The thickness is shown relative
to the photon attenuation length, this length is larger for the (400) plane because of the higher
photon energy. For the same absolute thickness, the angular yield with the (220) plane
is larger, as seen in Table \ref{table: FAST_yields_present} for a single thickness but 
the angular yield for the same relative crystal thickness is higher with the (400) plane because 
the absolute thickness is larger. In both cases, the angular yield appears to saturate at a
thickness of about 1.2$L_a$. 
The spectral brilliance for the same relative crystal thickness is larger with the (2,2,0) 
plane because the average emittance over the crystal is smaller with a smaller absolute 
thickness. In both cases, the brilliance reaches a maximum around (0.2 - 0.3)$L_{a}$.
These plots show that the optimum crystal thickness depends on whether the photon yield or the
spectral brilliance is the object of interest. 


\subsection{New Goniometer} \label{subsec: newgon}

\begin{figure}
\centering
\includegraphics[scale=0.6]{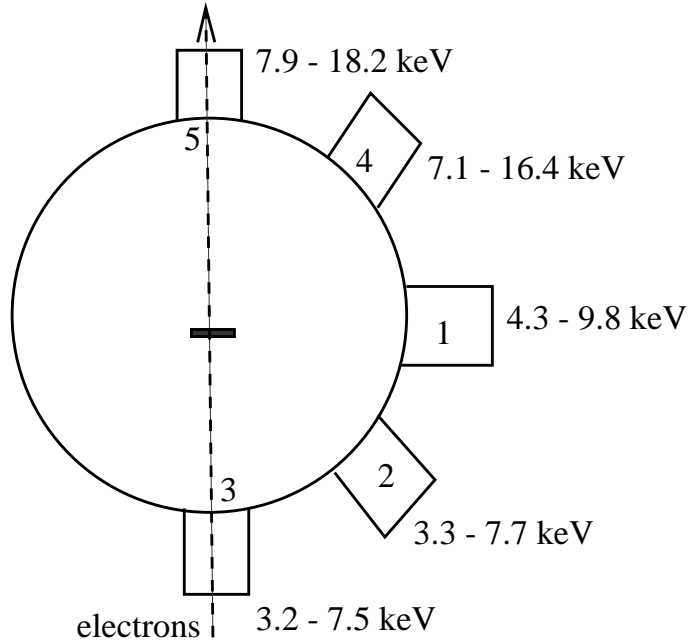}
\caption{Schematic of the new goniometer with the five labeled ports. The X-ray energies
expected from these ports with different planes, see Table 
\ref{table: newgon_EX}, are also
shown. Note: Ports 3 and 5 are not in the horizontal plane but are 
45$^{\circ}$ to the horizontal. There is an exit port under port 5 and in 
the horizontal plane but not shown here. }
\label{fig: newgonio}
\end{figure}

The initial set of experiments will be conducted with the goniometer described above with the
two ports. There is another goniometer under construction which will have a total of five ports
through which X-rays could be extracted. A schematic of this new goniometer is shown in 
Figure \ref{fig: newgonio}. 
These ports will offer the opportunity of generating PXR at different Bragg angles and hence
different energies, yields and spectral brilliance from those considered in the previous
subsection. 

The angle of the port determines the direction of the photon emission and hence the vector
$\vec{\Om}$. We write the unit vector along the detector axis as
$\hat{\Om}= (\om_1, \om_2, \om_3)/\sqrt{\om_1^2+ \om_2^2 +\om_3^2}$. 
If the PXR plane has Miller indices $(h,k,l)$, the unit normal $\hat{g}$ can be one of
$\pm (h,k,l)/\sqrt{h^2+k^2+l^2}$. The conditions for reflection require that 
both 
$\hat{g}\cdot(-\hat{v}) \ge 0$ and $\hat{g}\cdot \hat{\Om} \ge 0$.
The first of these results in choosing the sign of $\hat{g}$ while
the second determines if a chosen detector angle is suitable for observing
reflected photons. Together the two conditions ensure that the incident beam
and the reflected photons are on the same side of the reflecting plane. 
For a given plane with a normal $\hat{g}$, the Bragg angle is determined from
$\theta_B=\arcsin[-\hat{v}\cdot \hat{g}]$.
For the goniometer ports shown in Figure \ref{fig: newgonio},
the angle vectors to the detector locations are
\beqrs
\hat{\Om}_1 & = & (1,0,0), \;\;\; \hat{\Om}_2 = (\sin(13\pi/18),0,\cos(13\pi/18)),
\;\;\; \hat{\Om}_3 = (0,\frac{1}{\sqrt{2}},-\frac{1}{\sqrt{2}}), \\
\hat{\Om}_4 & = &  (\sin(5\pi/18),0,\cos(5\pi/18)), \;\;\;
\hat{\Om}_5 = (0,\frac{1}{\sqrt{2}},-\frac{1}{\sqrt{2}})
\eeqrs
Energies and angular yields at the different ports from different PXR planes are shown in 
Table \ref{table: newgon_EX}. 
\begin{table}
\bec
\btable{|c||c|c|c|c|} \hline
Port & \multicolumn{2}{|c|}{Plane 111} & \multicolumn{2}{|c|}{Plane 220} \\
\cline{2-5}
   &  Energy  & Ang. Yield & Energy  & Ang. Yield   \\ 
   &  [keV]  & [$\times 10^{-5}$ phot/e$^-$-sr ] & [keV] & [$\times 10^{-5}$ phot/e$^-$-sr ] \\
\hline
1 &   4.3 & 0.037  & 6.9  & 9.9     \\
2 &   3.3  & 3.4$\times 10^{-5}$   & 5.4  & 1.2  \\
3 &   3.2  & 1.5$\times 10^{-5}$   & 5.3  & 0.096  \\
4 &   7.1  &  76.9   & 11.6 & 90.5   \\
5 &   7.9  & 142  & 12.8 & 108.3 \\
\hline
Port & \multicolumn{2}{|c|}{Plane 311} & \multicolumn{2}{|c|}{Plane 400} \\
\cline{2-5}
   &  Energy [keV] & Ang. Yield & Energy [keV] & Ang. Yield   \\
   &  [keV]  & [$\times 10^{-5}$ phot/e$^-$-sr ] & [keV] & [$\times 10^{-5}$ phot/e$^-$-sr ] \\
\hline
1 & 8.1 & 5.3   & 9.8  & 8.8      \\
2 & 6.4 & 1.6   & 7.7  & 4.8     \\
3 & 6.2 & 1.4   & 7.5  & 4.6      \\
4 & 13.6 & 26.5   & 16.4 & 34.7       \\
5 & 15.1 & 41.3  & 18.2 & 40.8      \\
\hline
\etable
\eec
\caption{X-ray energies at the different ports for different PXR reflection
planes. Bragg geometry is assumed in all cases. }
\label{table: newgon_EX}
\end{table}

\begin{table}
\bec
\btable{|c||c|c|c||c|c|c|} \hline
Port &  \multicolumn{3}{|c|}{Plane 111} & \multicolumn{3}{|c|}{Plane 220} \\ \cline{2-7}
 &  Energy & $t/L_a$ & Sp. Br. [$\times 10^8$] & Energy & $t/L_a$ & Sp. Br. [$\times 10^8$] \\
\hline
 4 & 7.1  &  0.36 &  5.5 & 11.6  &  0.09  &  2.2 \\
 5 &  7.9 &  0.26  & 5.7  & 12.8   &  0.07 &  2.1 \\
\hline
Port &  \multicolumn{3}{|c|}{Plane 311} & \multicolumn{3}{|c|}{Plane 400} \\  \cline{2-7}
 &  Energy & $t/L_a$ & Sp. Br. [$\times 10^8$] & Energy & $t/L_a$ & Sp. Br. [$\times 10^8$] \\
\hline
4 & 13.6 & 0.06  &  0.59 & 16.4 &  0.04 &  0.59  \\
5 & 15.1 &  0.05 &  0.57 & 18.2 & 0.03  & 0.56 \\
\hline
\etable
\eec
\caption{PXR photon energy, crystal thickness $t$ relative to attenuation length $L_a$, and 
spectral brilliance (Sp. Br.) in units of photons/s-(mm-mrad)$^2$- 0.1\% BW] at ports 4 and 5 for 
different planes. The crystal thickness was 0.168mm in each case. }
\label{table: SpBr_NG}
\end{table}
Table \ref{table: SpBr_NG} shows the spectral brilliance expected from ports 4 and 5, the ones
corresponding to the smallest Bragg angles and highest yields. 
With the crystal thickness kept constant at 0.168 mm, the spectral brilliance is highest
for the (111) plane at these ports. The X-ray energy is
between 7-8 keV and the ratio of the crystal thickness to attenuation length is close to 
the optimal value of around 0.2, seen in Figure \ref{fig: yield_sr_Brilliance_LC}. 
For the higher order planes, the PXR energy increases but the spectral brilliance
decreases. Especially for planes (3,1,1) and (4,0,0) the brilliance drops by an order of 
magnitude compared to the (1,1,1) plane. This is partly due to the small relative thickness
and increasing the crystal thickness would also increase the brilliance, but not significantly.
The choice of plane would then be determined by whether higher energy or higher brilliance is
more desirable. A higher energy beam, e.g. 100MeV, would increase the yield and also the
brilliance because the emittance growth due to multiple scattering would also be smaller.

\subsection{PXR while channeling}

It was pointed out \cite{Yabuki_2001} that if a beam is channeled within a 
crystal and emits channeling radiation, it may also emit PXR emission from
reflection off complementary planes which intersect the channeling planes.
It has subsequently been observed at the SAGA light source linac with 255 MeV electron beams
\cite{Takabayashi_2013}. 
Here we consider the prospect of detecting PXR emission under channeling
conditions while using the present goniometer. As mentioned previously, this goniometer has
a second port at 90$^{\circ}$ to the beam axis which could be used for detection of PXR. These 
requirements impose constraints on the possible PXR planes and the orientation of the
crystal which we now consider. 

\begin{figure}
\centering
\includegraphics[scale=0.75]{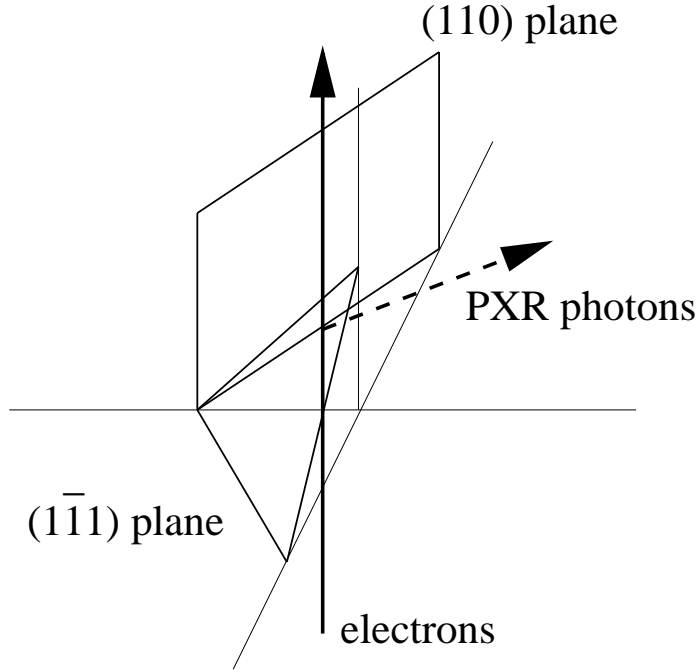}
\caption{Sketch of the (110) channeling plane, the $(1,\bar{1},1)$ plane which
is one of the possible PXR planes, the electron beam direction and the
direction of the PXR photons from this reflection plane.}
\label{fig: 110_111_planes}
\end{figure}
Choosing the $(1 1 0)$ plane as the channeling plane, the unit normal to this
plane is $\hat{m}=(1,1,0)/\sqrt{2}$. This vector must be normal to the velocity vector
which could be of the form $\hat{v} = (a,-a,c)/\sqrt{2a^2 + c^2}$
where $(a,c)$ are arbitrary real numbers. We will consider two choices below,
one with $(a,c)= (-1,0)$ and another with $(a,c) = (0,1)$. 
If the PXR plane is $(1 \bar{1} 1)$, the unit normal to the
PXR plane is $\hat{g}=(1,-1,1)/\sqrt{3}$ and the Bragg angle $\theta_B$ is
determined by the condition
$ \sin\theta_B = | \hat{g}\cdot \hat{v}| = |(2a+c)/(\sqrt{3(2a^2+c^2)})| $.
The choice $\hat{v}=(-1,1,0)/\sqrt{2}$ yields 
$\theta_B={\rm arcsin}(\sqrt{2/3})=54.7^{\circ}$ while
choosing the vector $\hat{v}=(0,0,1)$ yields 
$\theta_B={\rm arcsin}(\sqrt{1/3})= 35.3^{\circ}$. A smaller Bragg angle leads
to a higher PXR photon energy, so we choose $\hat{v}=(0,0,1)$. In practice, with a given electron
beam direction, these choices will correspond to different orientations of the crystal with 
respect to the beam velocity. For particles at other angles in 
the beam distribution, 
the requirement for channeling is that the velocity vector
has an angle smaller than the critical angle $\theta_C$, i.e.
$\theta = {\rm arcsin}(\hat{m}\cdot \hat{v}) < \theta_C $.

If we assume that the crystal has been cut so that the surface is parallel
to the channeling plane, then the unit normal to the surface, defined as
$\hat{n}$  (see Fig. \ref{fig: geometries})
is identical to the 
vector $\hat{m}$ defined above, i.e. $\hat{n} = \hat{m}= (1,1,0)/\sqrt{2}$.

\subsubsection{Crystal rotation}

If the crystal is aligned with so that the channeling plane is parallel to
the velocity vector, then the Bragg angle with a chosen PXR plane may not be 
appropriate to observe the PXR photons at 90 degrees to the beam. However, a
rotation of the crystal about an axis orthogonal to the channeling plane can
create the desired Bragg angle while maintaining the channeling condition.
If the rotation matrix about the normal to the channeling plane $\hat{m}$ is written as $R(\psi)$ 
where $\psi$ is the rotation angle, then the rotated normal to the PXR plane is
$\hat{g}_R = R(\psi)\hat{g}$. The requirement on this angle $\psi$
is that the velocity has the desired Bragg angle with the new normal $g_R$, 
i.e.
\beq
\sin\theta_B = \hat{g}_R \cdot (-\hat{v}) \equiv R(\psi)\hat{g} \cdot (-\hat{v})
\label{eq: angle_psi}
\eeq
The rotation matrix about an arbitrarily chosen unit vector
 $\hat{\mu}= (\mu_x, \mu_y, \mu_z)$ is the so-called Rodrigues matrix \cite{Rodrigues} given by
\beq
R(\psi) = \left[ \begin{array}{ccc} \cos\psi + \mu_x^2 S & -\mu_z\sin\psi +
\mu_x \mu_y S & \mu_y\sin\psi + \mu_x\mu_z S \\
\mu_z\sin\psi + \mu_x \mu_y S &  \cos\psi + \mu_y^2 S & -\mu_x\sin\psi + 
\mu_y\mu_z S \\ -\mu_y\sin\psi + \mu_x\mu_z S & \mu_x\sin\psi + \mu_y\mu_z S  &
 \cos\psi + \mu_z^2 S \end{array} \right]
\eeq
where $S = 2\sin^2(\psi/2)$. 
When the channeling plane is (110), the rotation axis is 
$\hat{\mu} = (1,1,0)/\sqrt{2}$.
The desired angle of rotation $\psi$ is found from Eq.(\ref{eq: angle_psi}) 
with $\theta_B=\pi/4$. 

Applying this to different possible PXR planes will ensure that the rotated light vector lies in 
the $(x,y)$ plane. A 
further rotation about the $z$ axis (i.e along the beam direction) may be
necessary to direct the light out along the $x$ axis. When the emitted light
as viewed at an angle 2$\theta_B$ to the beam, the reflected light vector 
prior to any rotations is $\hat{\Om} = \hat{v} - 2(\hat{g}.\hat{v})\hat{g}$
while that after rotation is $\hat{\Om}_R =
\hat{v} - 2(\hat{g}_R \cdot \hat{v})\hat{g}_R$.

As an example, consider the plane $(\bar{1}, 1, 1)$ for PXR production.
Choosing the normal to this plane as $\hat{g} = (1,-1,-1)/\sqrt{3}$, we have
\[ \hat{g}_R\cdot(-\hat{v}) = 
\frac{1}{\sqrt{3}}(\cos\psi + \sqrt{2}\sin\psi) =\frac{1}{\sqrt{2}} \Rarw
\psi =  9.74^{\circ}, \hat{\Om}_R  = \frac{1}{\sqrt{2}}(1, -1,0)
\]
In this case, a rotation by $\psi = 9.74^{\circ}$  about the normal to the
channeling plane and a further rotation by $\phi=45^{\circ}$ about the beam
axis will direct the PXR light out of the port along the positive $x$ axis.

Table \ref{table: PXRC_present} shows some of the low order planes, the angles of 
rotation $\psi$ and $\phi$, and the energies of the emitted X-rays into the 
detector. This table shows that the $(2,0,\bar{2})$ plane would be the
simplest as it does not require any rotation of the crystal while it is
oriented for channeling. Another reason for choosing this plane is that with rotation, the 
path length of the beam while channeling will be different compared to the unrotated case and 
could affect the channeling yield. 

\begin{table}
\bec
\btable{|c|c|c|c|} \hline
Plane &  Angle$\psi$[deg] & Angle $\phi$[deg] & $E_X$[keV] \\ \hline
            &    &    &   \\
$(\bar{1}, 1, 1)$ & 9.74 & 45 &  4.26 \\
$(1, \bar{1}, 1)$ & -9.74 & 135 & 4.26 \\
(2, 0. 2) & 0 & 180 & 6.95 \\
$(2,0,\bar{2})$ & 0 & 0 & 6.95 \\
$(0,2,\bar{2})$ & 0 & -90 & 6.95 \\
(0,0,4) & 45 & 135 & 9.83 \\
\hline
\etable
\eec
\caption{PXR planes, the angles of rotation $\psi$ about the normal to the
channeling plane (1,1,0) and $\phi$, angle of rotation about the beam direction
so that the PXR light is directed out of the port along the positive $x$ 
direction at 90 degrees to the beam axis. The energy of the PXR light is also
shown. Other low-order planes not shown are those for which no rotation will direct the
X-ray beam in the desired direction.}
\label{table: PXRC_present}
\end{table}

The angle $\zeta$ between the unit normal $\hat{n}$ to the surface and the unit normal
$\hat{g}$  to the channeling
plane will change depending on the channeling plane chosen. We assume here as above 
that the crystal is cut parallel to the channeling plane so that 
$\hat{n} = (1,1,0)/\sqrt{2}$. The angle between the two is given by 
$\zeta = \arccos(\hat{n}\cdot\hat{g})$, thus for the $(2,0,\bar{2})$ PXR plane, 
this angle is 60$^{\circ}$.

\subsubsection{Yield and spectral brilliance}

The above analysis has shown that with the crystal oriented for channeling
along the (1,1,0) plane, PXR emission from the (2,0,$\bar{2}$) plane can be
obtained at 90$^{\circ}$ to the beam direction without any crystal rotation.
In the subsequent discussion, we will primarily consider PXR from this plane.

When beam energies are under 100 MeV, the channeling radiation spectrum 
consists of a few well defined lines and is best understood as arising from
transitions between bound states of the transverse potential. 
We consider planar channeling and the $x$ direction to be orthogonal to the
channeling plane. The quantum mechanical states are then found from solutions 
of the one dimensional Schrodinger equation,
\beq
[-\frac{\hbar^2}{2m_e \gm} \frac{\del^2}{\del x^2} + V(x)]\psi(x) = E_{\perp}\psi(x)
\eeq
Here $V(x)$ is the one dimensional continuum potential obtained by averaging 
the
three dimensional atomic potential along the orthogonal directions $(y,z)$.

Parametric X-rays emitted under channeling conditions (PXRC) are considered to
be those in which the electrons stay within the same transverse energy band, 
i.e with only intra-band transitions. PXR emitted while electrons 
transition between different energy bands is labeled as diffracted channeling
radiation (DCR) and has apparently not yet been experimentally observed. 
We will only consider the yield from PXRC here.

The angular spectrum of PXRC is related to that of PXR by 
\cite{Korotchenko_12}
\beq
\frac{d^2 N}{d\theta_x\theta_y }|_{PXRC} = 
\frac{d^2 N}{d\theta_x\theta_y }|_{PXR} \sum_n^{N_B} P_n|F_{nn}|^2
\eeq
where the sum over the states $n$ ranges over the total number of bound states
$N_B$. 
Here we also include the effects of a finite beam divergence, so we define
$P_n$ the initial probability of occupation of state $n$ by 
averaging over the beam divergence. It is given by
\beq
\lan P_n \ran = \frac{1}{\sqrt{2\pi}\sg_x^{'}}\frac{1}{d_p}
\lan \int d\phi_{x,0} \exp[-\frac{\phi_{x,0}^2}{2(\sg_x^{'})^2}]
|\int_{-d_p/2}^{d_p/2} \exp[-i k \phi_{x,0} x]
\psi_{n, K_C}(x) dx |^2  \ran_{K_C}
\eeq
Here $\sg_x^{'}$ is the beam divergence in the channeling plane, 
$d_p$ is the inter-planar separation, $k$ is the initial momentum 
wavenumber of the incident particle,
$\phi_{x,0}$ is the angle of incidence with respect to the channeling plane
and $\psi_{n, K_C}$ is the wave function in the $n$th state with
transverse wavenumber $K_C$ where $-g/2 \le K_C \le g_2$. Since this 
wave function
also depends upon the band wavenumber $K_C$ in the Brillouin zone, the average
 $\lan \ran_{K_C}$ on the right hand side represents an average over the 
wavenumbers in the Brillouin zone.

The form factor $F_{nn}$ describes the impact of the channeling wave functions on the PXR yield
and is defined as \cite{Yabuki_2001}
\[  F_{nn} = \lan \psi_n | \exp[- i k_x x] |\psi_n \ran \]
where $k_x$ is the transverse component of the photon wave vector. 
We can therefore define the averaged form factor squared as
\beq
\lan |F_{nn}(\theta_x)|^2 \ran = \frac{1}{d_p^2}\lan |\int_{-d_p/2}^{d_p/2} \psi_{n,K_C}^*(x)
\exp[-i \frac{\om_B \theta_x x}{c}] \psi_{n, K_C}(x) dx |^2 \ran_{K_C}
\eeq
Here $\theta_x$ is the angle of photon emission in the horizontal plane with 
respect to the direction of specular reflection and we average over the 
Brillouin zone as before. 

The quantum mechanical calculations were done with a Mathematica notebook
used at the ELBE facility to study channeling \cite{Azadegan_13} and
significantly corrected and modified for use at FAST, as described in \cite{Sen_14}. 
The transverse function is expanded in a Fourier series using the lattice
periodicity as
\beq
V(x) = \sum_{n=-\infty}^{\infty} V_n \exp[i n g x]
\eeq
The Fourier coefficients $V_n$ are obtained from the Doyle-Turner coefficients
$(a^{DT}, b^{DT})$ \cite{Doyle-Turner} as 
\beq
V_n = -\frac{2\pi}{V_c}a_0^2 (\frac{e^2}{a_0})e^{-M(\vec{g})}
\sum_j e^{i \vec{g}\cdot \vec{r}_j} \sum_{i=1}^6 a_i^{DT}
\exp[-\frac{b_i^{DT}}{16\pi^2}(n g)^2]
\label{eq: V_n}
\eeq
Here $V_c$ is the volume of the 
unit cell, $a_0$ is the Bohr radius, $\vec{r}_j$ are the coordinates of the $j$th
atom in the unit cell and $M(\vec{g}) = \half g^2 \lan u^2  \ran$ is the 
Debye-Waller factor (mentioned earlier with Eq.(\ref{eq: intensity_spect}) in 
Section \ref{sec: spectrum}) describing thermal 
vibrations with mean squared amplitude $\lan u^2\ran$, assumed to be the same for all atoms. 
The wave functions are found by first expanding them in a series of Bloch 
functions and then solving the resulting matrix eigenvalue problem, see e.g
\cite{Azadegan_PRB}. 

The wave functions are then subsequently used in calculating the 
correction factors defined above. The form factor 
$\lan |F_{nn}(0)|^2 \ran = 1$ and we find that in the range 
$-3/\gm \le \theta_x \le 3/\gm$, $\lan |F_{nn}(\theta_x)|^2 \ran \approx 1$
to within 0.2\% in all cases.
The significant correction for the PXRC yield is due to the initial 
probability of occupation $\lan P_n \ran$ which is determined by the 
beam divergence. This population of the bound states decreases as the
beam divergence approaches the critical angle of channeling which at the
FAST energy of 50MeV is about 0.98mrad. The
correction $\dl$ is defined as the relative difference between the PXR and PXRC yields, hence as
$\dl = 1 - \sum_n^{N_B} \lan  P_n \ran \lan |F_{nn}|^2 \ran$. This correction factor increases 
with beam divergence and has the values 0.02, 0.09 and 0.58 at
beam divergences of 0.01mrad, 0.1mrad and 1mrad respectively.

\begin{figure}
\centering
\includegraphics[scale=0.55]{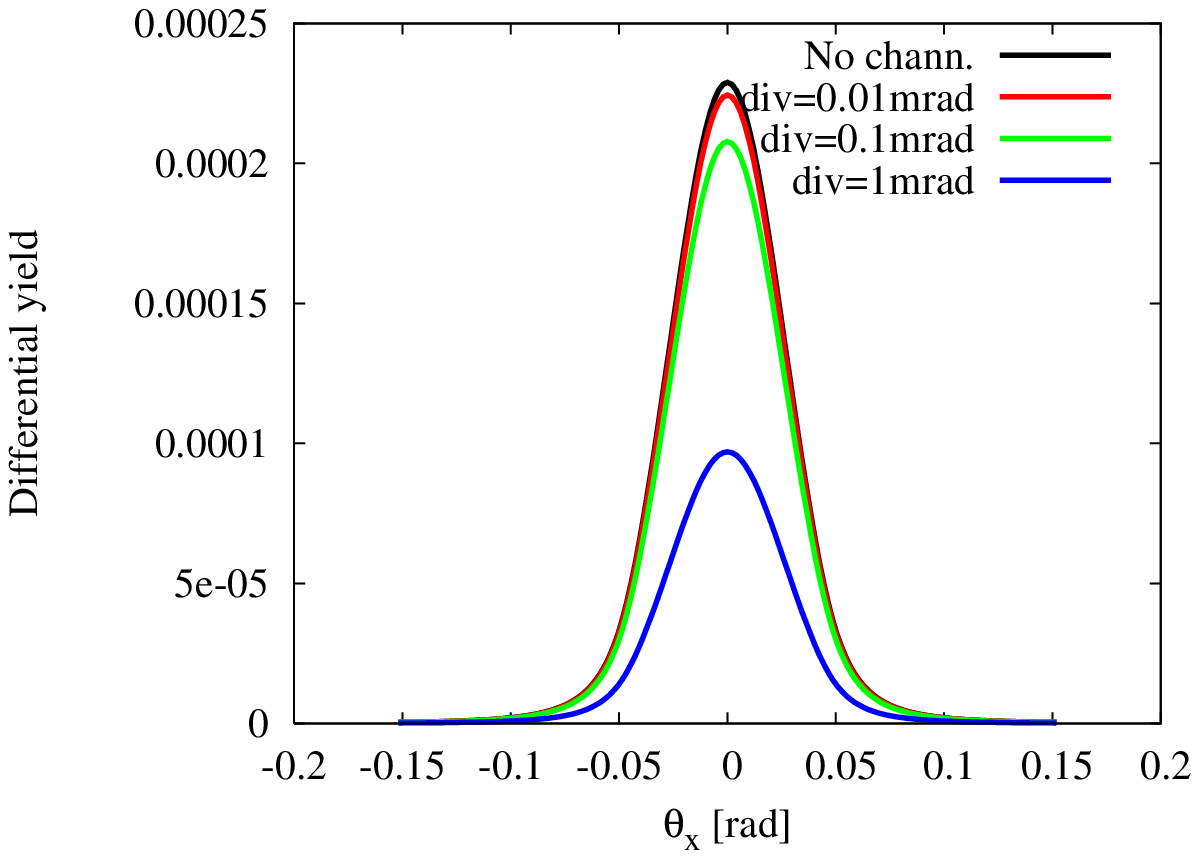}
\includegraphics[scale=0.55]{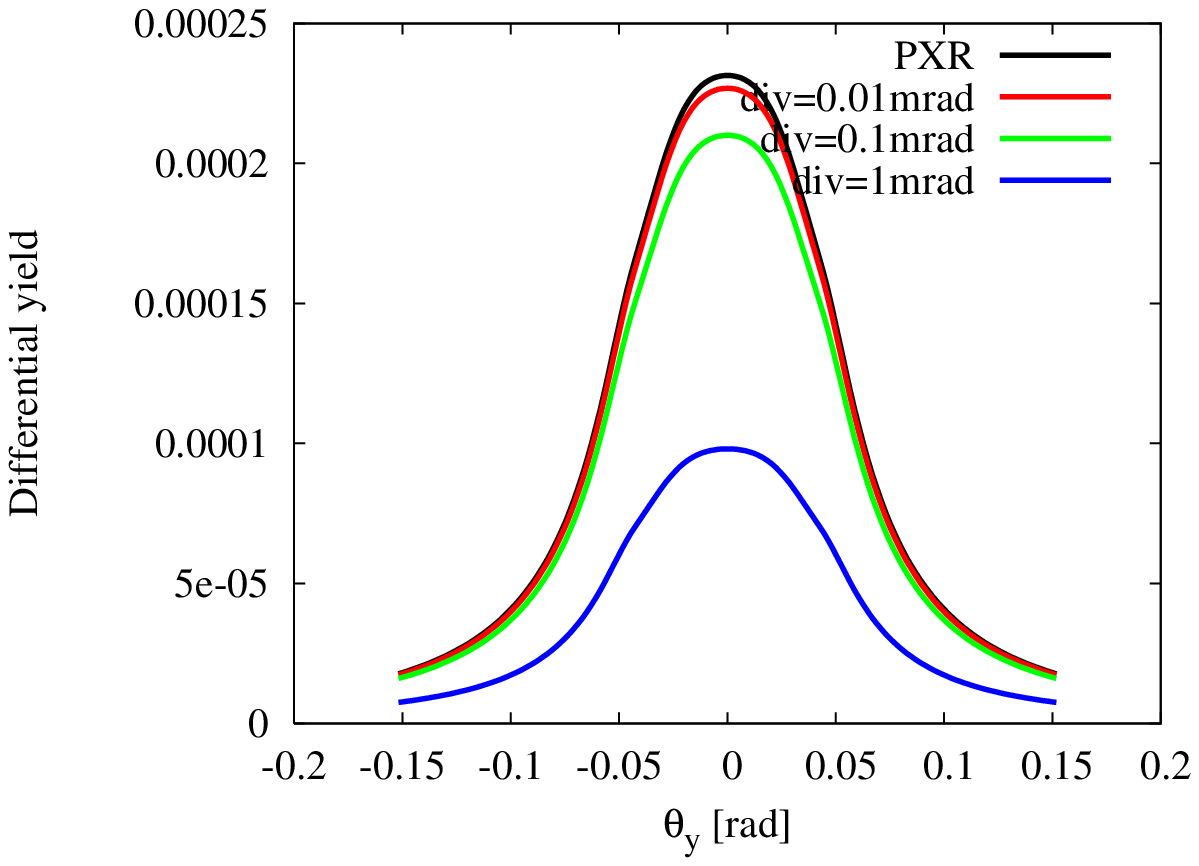}
\caption{The differential yield as a function of the photon emission angle
$\theta_x$ (left) and $\theta_y$ (right) without channeling  and with channeling 
for three values of the initial beam divergence. The channeling plane was (1,1,0) and the PXR 
plane was (2,0,$\bar{2}$). Convolution broadening was included in all cases. }
\label{fig: pxr_c_compyields}
\end{figure}

\begin{table}
\bec
\btable{|c||c|c|c|c||c|c||} \hline
Plane &  Divergence & $\Dl E$ & \multicolumn{2}{|c||}{Yield} & \multicolumn{2}{|c||}{Spectral Brilliance} \\
 & mrad & eV & \multicolumn{2}{|c||}{$\times 10^{-4}$ photons/(e$^{-}$-sr)} & 
 \multicolumn{2}{|c||}{photons/(mm-mrad)$^2$-0.1\% BW} \\ \cline{4-7}
 & &  &  PXR &  PXRC  &    PXR &  PXRC  \\ \hline
(2,0,$\bar{2}$) & 0.01 & 93 & 0.40   & 0.38  & 35.9  & 35.1   \\
 &  0.1 & 93 &  0.40 & 0.35 & 2.1$\times 10^5$  &   1.9$\times 10^5$  \\
 &  1.0 & 93 & 0.40  & 0.16  & 6.5$\times 10^8$ & 1.8$\times 10^8$  \\
\hline
(0,0,4) &  0.01 & 131 & 0.89   & 0.85  & 21.3  & 20.8  \\
 &  0.1 & 131 &  0.89 & 0.78   & 1.27 $\times 10^5$ & 1.14 $\times 10^5$  \\
 & 1.0 & 131 &  &  0.37 &  4.25 $\times 10^8$ & 1.2$\times 10^8$ \\
\hline
\etable
\eec
\caption{Comparison of angular yields, photon flux and spectral brilliance
without and with channeling for three values of the initial beam
divergence and for two PXR planes. The channeling plane was fixed at (1,1,0).
Angle $\zeta=60^{\circ}$ for the 
(2,0,$\bar{2}$) plane, and $\zeta=90^{\circ}$ (Laue geometry) for the (0,0,4) plane.
The energy width is affected by the PXR  plane but not by
the initial beam divergence.}
\label{table: PXRC_yields}
\end{table}
Figure \ref{fig: pxr_c_compyields} shows the angular spectra from PXRC for the three
divergences mentioned and compared with the spectrum from PXR without channeling. 
Table \ref{table: PXRC_yields} shows that the angular yield decreases with beam divergence, 
because the  fraction of particles in the bound states decreases with increasing 
divergence. Since the spectral brilliance is inversely proportional to the square of the
spot size and assuming the emittance is conserved, a larger beam divergence
implies a smaller spot size. The increase in brilliance with the smaller spot 
size dominates the decrease due to channeling.

\subsection{Possible applications of PXR}

One of the primary foreseen uses of PXR is in phase 
contrast imaging (PCI) of low Z materials, specially biological samples. There are many PCI
methods, we plan to use the free space propagation method which does not need any additional 
optical elements. The principle of PCI is that phase shifts undergone by hard X-rays when
traversing low Z samples are orders of magnitude larger than absorption effects. The phase shift
changes the complex amplitude of the wave field and hence causes intensity modulations when
undergoing interference with X-rays that have not passed through the sample. The quality of the
image with the free space propagation method depends on the source size, the geometric
magnification and the resolution of the detector. Th special feature of FAST that makes it
suitable for PCI is the very low emittance of the electron beam, and hence the small X-ray 
source size after the crystal. With electron emittances around 100nm using a
conventional photocathode, beam sizes at the crystal around 1-5 $\mu$m can be achieved
\cite{Piot_12}. This compares very favorably with recent PCI experiments  using
PXR \cite{Hayakawa} which had beam sizes of 0.5 - 2 mm (FWHM) at the target. With the use of 
field emitter
nanotips as the cathode, the beam emittance could be improved another order of magnitude 
\cite{Piot_14} implying a further improvement in image resolution. The use of the new 
goniometer, discussed in Section \ref{subsec: newgon} will enable imaging at multiple X-ray 
energies.

Finally, we mention a proposal \cite{Thangaraj} to generate short electron bunches at FAST
using a slit mask placed in the middle of the bunch compressor chicane. Sub-picosecond
electron bunches could be produced without the need of an undulator or an additional complex
laser system. Due to scattering in the mask, the final beam intensity will only be about 10\%
of the initial intensity. However, starting from initial intensities of 1-3 nC, the final
bunch intensities within the bunch train will be low enough (about 20 pC) for PXR generation.
The resulting sub-picosecond X-ray pulses will have a higher peak brilliance and could be used 
for time resolved X-ray studies in materials science, chemistry and biology.

\section{Conclusions}

In this paper we considered the prospect of generating PXR using diamond crystals with
50 MeV electron beams from the photoinjector at the FAST facility at Fermilab. We
revisited calculations of the energy width from both geometric and multiple scattering. 
Comparisons with earlier experiments were found to yield reasonable agreement
The PXR spectrum model calculation was applied to the conditions at FAST.
Using the presently available goniometer restricts the Bragg angle to 45$^{\circ}$ but allows
a clear separation from the electron beam. 
The PXR energies from the planes studied fall in the range 4 - 10 keV with a spectral
energy widths of $\sim 1$\%. With a diamond crystal thickness of 168$\mu$m, maximum angular yields
are about 10$^{-4}$ photons/e$^-$-sr, taking into account attenuation in air from the crystal
to the detector. Using crystals of different thicknesses, 
the PXR yield was found to saturate at a thickness of about 1.2$L_a$,
the  photon attenuation length  at two different energies.
The spectral brilliance on the other hand attained a maximum value at around 0.2$L_a$ for
these same energies. This is mainly because the emittance growth over larger thicknesses 
reduces the brilliance more than the yield increases it. 
Next, PXR emission with a new goniometer with five possible ports was studied. The range
of energies now spans 3 - 18 keV. The spectral brilliance, around 10$^8$
photons/(mm-mrad)$^2$-0.1\% BW, is reached with a fixed crystal thickness which is
less than 0.2$L_a$ for most energies, so higher brilliance is feasible with thicker crystals. 
Use of this goniometer would open up the possibility of extracting PXR at multiple energies 
simultaneously. 
Finally, PXRC or PXR under channeling conditions was studied and the yield with quantum corrections
from channeling was calculated for three beam divergences. While the reduction in PXR yield
during channeling was smallest for the lowest divergence, higher brilliance favors the smallest
beam spot size or equivalently the largest divergence under conditions of equal emittances.
This PXRC emission makes possible simultaneous X-ray emission from channeling and PXR at 90 
degrees to each other. 
The brilliance of PXR appears to be sufficient for phase contrast imaging and the FAST 
facility with (10 - 100) nm scale electron emittances should enable imaging with very good 
resolution.

\vspace{2em}

\noi {\bf \large Acknowledgments} \newline
We thank the Lee Teng undergraduate internship program at Fermilab which awarded 
T. Seiss a summer internship in 2014. Fermilab is operated by Fermi
Research Alliance under DOE Contract No. DE-AC02-07-CH11359.

\end{document}